\documentclass[aps,pra,10pt,a4paper,notitlepage,nofootinbib, superscriptaddress,twocolumn]{revtex4-2}

\usepackage{bm,graphicx,textcomp,amssymb,amsmath,dcolumn,color}
\usepackage{epstopdf}

\newcommand{\Tr}{\operatorname{Tr}}

\renewcommand{\rm}{\mathrm}
\newcommand{\mbf}{\mathbf}

\usepackage[dvipsnames]{xcolor}

\newcommand{\ket}[1]{\left|#1\right>}

\newcommand{\nn}{\nonumber\\}

\newcommand{\bea}{\begin{eqnarray}}
\newcommand{\ea}{\end{eqnarray}}
\newcommand{\eea}{\end{eqnarray}}
\newcommand{\ord}{\,{\cal O}}

\begin{document}

\title{Decay of quantum sensitivity due to three-body loss in Bose-Einstein condensates}

\author{Dennis R\"atzel}
\email{dennis.raetzel@physik.hu-berlin.de}
\affiliation{Institut f\"ur Physik, Humboldt-Universit\"at zu Berlin, 
Newtonstra\ss e 15, 12489 Berlin, Germany}

\author{Ralf~Sch\"utzhold}
\affiliation{Helmholtz-Zentrum Dresden-Rossendorf, Bautzner Landstra{\ss}e 400, 01328 Dresden, Germany,}
\affiliation{Institut f\"ur Theoretische Physik, Technische Universit\"at Dresden, 01062 Dresden, Germany.}

\date{\today}

\begin{abstract}
In view of the coherent properties of a large number of atoms, Bose-Einstein Condensates (BECs) have a 
high potential for sensing applications. 
Several proposals have been put forward to use collective excitations such as phonons in BECs for quantum 
enhanced sensing in quantum metrology. 
However, the associated highly non-classical states tend to be very vulnerable to decoherence. 
In this article, we investigate the effect of decoherence due to the omnipresent process of three-body 
loss in BECs.  
We find strong restrictions for a wide range of parameters and we discuss possibilities to limit these 
restrictions.
\end{abstract}


\maketitle

\section{Introduction} 

After their first experimental realization \cite{Davis:1995Bose,anderson_observation_1995} in 1995, 
Bose-Einstein condensates (BECs) in ultra-cold atomic vapor are now routinely created and manipulated 
in many laboratories. 
Since then, experimentalists have gained a significantly higher level of control over BECs through 
technological and methodological advancements, including 
low-temperature records in the sub-nK regime \cite{leanhardt_cooling_2003}, 
creation and manipulation on an atom-chip \cite{schumm_matter-wave_2005,wildermuth_microscopic_2005}, 
and sending BECs to space \cite{lachmann_creating_2018}, 
allowing fundamental research in a microgravity environment. 

In state-of-the-art technology, BECs are used for high precision measurements of forces 
\cite{Muentinga:2013int,Altin_2013,hamilton_atom-interferometry_2015,Hardman:2016sim,
Abend:2016atom,haslinger_attractive_2018} by means of matter-wave interferometry, 
where the wave function of each atom is split into two wave packets that are sent 
on different paths and then brought into interference. 

For optical interferometry, it is well known that one may enhance the sensitivity by employing 
non-classical (e.g., squeezed) states, which has been successfully implemented in the gravitational 
wave detectors of the LIGO-Virgo collaboration \cite{aasi_enhanced_2013,Acernese:2019incr,Tse:2019quant},
for example. 
More generally, quantum metrology refers to the exploitation of quantum properties (such as entanglement)
in order to gain significantly higher sensitivities for measurement technologies 
\cite{Giovannetti2006quant}. 
Quantum enhanced sensing has also been proposed for matter-wave interferometry 
\cite{Salvi:2018squee} and other sensing applications of cold atomic systems \cite{Pezze:2018quant}.

For cold atoms however,
the strict conservation of their 
total number 
poses restrictions on the available phase space of quantum superpositions. 
Thus, it might be advantageous to employ the quantum states of collective oscillations such as 
phonon modes in the BEC.

First studies of collective oscillations in BECs were already performed in early experiments 
\cite{Jin:1996coll,Mewes:1996coll,Stamper-Kurn:1998coll}. 
Highly excited quasi-particle states can be created with light pulses \cite{Katz:2004high} 
and periodic modulations of the trap potential \cite{Jaskula:2012acoustic,Michael2019from}. 
Measurement methods include self-interference of the Bose gas after release from the trap 
denoted as heterodyning \cite{Katz:2004high} or time-of-flight measurements and in-situ 
phase contrast imaging \cite{Stamper-Kurn:1998coll,Schley:2013planck}. 
A specific example of the utilization of collective oscillations in BECs for sensing was 
the measurement of the thermal Casimir-Polder force presented in 
\cite{Obrecht:2007meas,Antezza:2004effect}.   

Exploring the potential of BECs for sensing applications further, it has been proposed to use 
collective oscillations in BECs to measure the effect of space-time curvature on entanglement 
\cite{Bruschi_2014,Matos2015,howl_gravity_2018}, 
for high-precision gravity sensing \cite{ratzel_dynamical_2018,bravo2020phononic}, 
for detecting gravitational waves
\cite{Sabin:2014gravwave,sabin_thermal_2016,Schuetzhold:2018int,Robbins_2019,robbins2021detection}
and for testing gravitationally induced collapse models \cite{Howl:2019expl}. 

In order to detect extremely small effects (such as the gravitational phenomena mentioned above),
many proposals for sensing with collective oscillations in BECs rely on elements of quantum sensing 
employing specific quantum states. 
However, these highly non-classical (e.g., squeezed) states are typically also quite prone 
to noise and decoherence \cite{escher_general_2011,Pezze:2018quant}.
In this work, we study the effects of decoherence caused by the omnipresent process of three-body 
loss in BECs. 
In contrast to other decoherence channels such as Landau or Beliaev damping (see, e.g., \cite{Howl_2017}), 
this mechanism is not suppressed when going to ultra-low temperatures (as for Landau damping) 
or energies (as for Beliaev damping).

\section{Three-body loss}\label{sec:three-body}

Bose-Einstein condensates of ultra-cold atomic vapor are not the true ground states, which would be solid. 
They are meta-stable states which can be rather long lived because two atoms alone cannot bind and form a 
molecule as energy and momentum conservation forbid them to dispose of the released binding energy. 
However, if a third atom is close by (i.e., within the short interaction range) and carries away the 
excess energy and momentum, this recombination process can occur.   
Since the energy scales (set by the binding energy) are typically much larger than the characteristic 
energy scales of the ultra-cold BEC, effectively all three atoms are lost from the BEC in such a process.
Furthermore, the associated length (interaction range) and time scales are much 
shorter than those of a BEC, such that one may approximate
the recombination process as local in space and time. 
Thus, we use the Born-Markov approximation 
and describe these three-body loss processes via a Lindblad master equation ($\hbar=1$), 
see also \cite{Jack:2002dec}
\bea\label{eq:Lindblad}
\frac{d\hat\varrho}{dt}
&=&
-i\left[\hat H_0 ,\hat\varrho\right]
+
\Gamma\int d^3r\,
\hat\Psi^3(\mbf{r})\hat\varrho [\hat\Psi^\dagger(\mbf{r})]^3
\nn
&&
-\frac{\Gamma}{2}\int d^3r
\left\{
[\hat\Psi^\dagger(\mbf{r})]^3 \hat\Psi^3(\mbf{r}),\hat\varrho
\right\}\,.
\ea
Here $\hat H_0$ is the usual Hamiltonian governing the undisturbed dynamics of the BEC and 
$\hat\varrho$ denotes its density matrix.
The Lindblad jump operators are given by third powers of the field operators $\hat\Psi(\mbf{r})$
in second quantization, each one corresponding to the annihilation of an atom at position $\mbf{r}$.
Finally, $\Gamma$ is the bare loss rate. 

In order to clearly distinguish different effects, we assume scale separation,
i.e., the characteristic length scales (e.g., size) of the condensate are supposed to be much larger 
than the wavelengths of the phonon modes under consideration -- which, in turn, should be much longer
than the typical length scales of the atomic interactions and the three-body losses.

\subsection{Mean-field approximation}\label{sec:mean-field}

As usual, we split the atomic field operator $\hat\Psi(\mbf{r})$ into the macroscopically occupied mode 
$\hat{a}_0$ described by the condensate wave-function $\psi_{\rm c}(\mbf{r})$ plus the field operator 
$\hat\chi(\mbf{r})$ for all other modes 
\bea\label{eq:split}
\hat\Psi(\mbf{r})
\approx
\psi_{\rm c}(\mbf{r})\hat a_0 
+ \hat\chi(\mbf{r})
\,.
\ea
This is followed by the Bogoliubov approximation, where the field $\hat\chi(\mbf{r})$ is treated as a 
small perturbation (in view of the large occupation of the condensate mode). 

Inserting split~(\ref{eq:split}) into the Lindblad equation~(\ref{eq:Lindblad}), the zeroth order 
in $\hat\chi(\mbf{r})$ yields the decay of expectation value of the number of atoms in the condensate 
\bea
\nonumber 
\frac{dN_{\rm c}}{dt} 
&=& 
\frac{d}{dt} \langle\hat a_0^\dagger\hat a_0\rangle 
\approx
- 3\Gamma\int d^3r\,|\psi_{\rm c}(\mbf{r})|^6 
\langle 
 (\hat a_0^{\dagger})^3 \hat a_0^3 
\rangle 
\\
&\approx& - 3\Gamma N_{\rm c}^3 \int d^3r\,|\psi_{\rm c}(\mbf{r})|^6
\,,
\ea
where we have used the standard approximation of the condensate as a coherent state in the last step. 
This can be reformulated in terms of the condensate density 
$\rho_{\rm c}(\mbf{r}) = N_{\rm c} |\psi_{\rm c}(\mbf{r})|^2$ 
which then leads to
\bea\label{eq:BECdecay}
\frac{d}{dt} N_{\rm c} = \frac{d}{dt} \int d^3r\, \rho_{\rm c}(\mbf{r}) 
&\approx& 
- D \int d^3r\,\rho_{\rm c}^3(\mbf{r})
\,,
\eea
where $D= 3\Gamma$. 
This is the usual equation describing three-body loss in BECs with decay constant $D$ 
(see, e.g., Section 5.4 of \cite{Pethick:2002bose} and in \cite{Norrie:2006thr,Moerdijk:1996dec}). 
For example, based on experiments with rubidium atoms, the corresponding decay constant was given as 
$D\sim 1.8\times 10^{-29}\,\rm{cm}^6\rm{s}^{-1}$ in \cite{Soding:1999three} and 
$D\sim 5.8\times 10^{-30}\,\rm{cm}^6\rm{s}^{-1}$ in \cite{Burt:1997coh} 
for different internal states of the atoms. 
In the following, we use the second result $D\sim 5.8\times 10^{-30}\,\rm{cm}^6\rm{s}^{-1}$ 
when we give numerical values for  rubidium 
BECs. 
In an experiment \cite{Takuso:2003spin} with ytterbium, a decay constant of 
$D \sim 4\times 10^{-30}\,\rm{cm}^6\rm{s}^{-1}$ was found.

As one may infer from the above equation~\eqref{eq:BECdecay}, three-body losses can be quite rare 
events in very dilute BECs, explaining the comparably long life-time of these meta-stable states. 
Still, they are always present and will have profound consequences.  
As we shall find below, the decoherence induced by three-body losses limits 
the accuracy of quantum enhanced sensing already on time scales much shorter than the 
life-time of the condensate.
On such short time scales, the depletion of the BEC according to Eq.~\eqref{eq:BECdecay} can be 
neglected, i.e., $N_{\rm c}\approx\rm{const}$, and we use this approximation in the following.

\subsection{Phonon modes}\label{sec:phonons}

The next-to-leading-order terms in $\hat\chi(\mbf{r})$ describe the dynamics of 
the state $\hat\varrho_{\chi}$ of
the Bogoliubov quasi-particle excitations (see also \cite{dziarmaga_bose_2003})
\bea
\label{eq:masterpho0}
\nonumber 
\frac{d\hat\varrho_{\chi}}{dt}
&=& 
-i\left[\hat H_0 ,\hat\varrho_{\chi}\right] + 9\Gamma N_{\rm c}^2 \int d^3r |\psi_{\rm c}(\mbf{r})|^4 
\times\\
&&\times\Big( \hat\chi(\mbf{r}) \hat\varrho_{\chi} \,\hat\chi^\dagger(\mbf{r})
-\frac12 \left\{ \hat\chi^\dagger(\mbf{r}) \hat\chi(\mbf{r})  ,\hat\varrho_{\chi} \right\} \Big) 
\,.
\ea
If the Hamiltonian $\hat H_0$ is bi-linear in the field operators $\hat\chi(\mbf{r})$ and 
$\hat\chi^\dagger(\mbf{r})$, this master equation preserves Gaussianity, i.e., 
maps initial Gaussian states $\hat\varrho_{\chi}$ to final Gaussian states.
This will become important later on, but at this point we do not restrict our considerations 
to Gaussian states.

Note that $\hat\chi^\dagger(\mbf{r})$ and $\hat\chi(\mbf{r})$ are not directly the 
creation and annihilation operators for the quasi-particle excitations, they are related via 
a Bogoliubov transformation 
\begin{equation}
\label{Bogoliubov-trafo}
\hat\chi(\mbf{r}) 
= 
\sum_{I} \hat{a}_I \phi_I(\mbf{r}) 
= 
\sum_{I} \left(u_I(\mbf{r})\hat b_I + v_I^*(\mbf{r})\hat b_I^\dagger\right)
\,.
\end{equation}
Here $\hat{a}_I$ are the original atomic annihilation operators for the modes $\phi_I(\mbf{r})$ 
orthogonal to the condensate wave-function $\psi_{\rm c}(\mbf{r})$ while $\hat b_I^\dagger$ and 
$\hat b_I$ are the quasi-particle creation and annihilation operators, respectively.  
The mode functions $u_I$ and $v_I$ fulfill the stationary Bogoliubov-de~Gennes equations 
and are ortho-normalized as
\begin{equation}\label{eq:normauvappend}
\int d^3r\, \left[u_I^*(\mbf{r}) u_J(\mbf{r}) - v_I^*(\mbf{r}) v_J(\mbf{r})\right] = \delta_{IJ}\,.
\end{equation}
As a result, the master equation~\eqref{eq:masterpho0} contains Lindblad operators $\hat b_I$
corresponding to cooling $\propto u_I$ as well as Lindblad operators $\hat b_I^\dagger$ describing 
heating $\propto v_I$,
see also \cite{dziarmaga_bose_2003}.
The perhaps somewhat surprising effect of heating (even at zero temperature)
can intuitively be understood in the following way:
Due to the small but finite interaction between the atoms, not all of them are in the macroscopically 
occupied mode $\hat a_0$ described by the condensate wave-function $\psi_{\rm c}(\mbf{r})$, a small 
fraction of them -- referred to as the quantum depletion -- 
is ``pushed'' to higher modes $\phi_I(\mbf{r})$, even in the quasi-particle ground state.   
Now, if one of the atoms involved in a three-body loss event belonged to the quantum depletion,
removing this atom would constitute a departure from the quasi-particle ground state, i.e., 
an excitation. 

\section{Decoherence}\label{sec:decoherence}

Now we are in the position to derive the decoherence of the phonon modes due to three-body loss.
Of course, to actually solve the master equation~\eqref{eq:masterpho0}, we have to specify the 
undisturbed Hamiltonian $\hat H_0$.
In the following, we discuss several examples. 

\subsection{Pure decay channel}\label{sec:decay}

As our first and most simple example, let us omit the undisturbed Hamiltonian $\hat H_0$ altogether.
As a further simplification, we focus on the decay channel, i.e., we only keep the Lindblad operators 
$\hat b_I$ and neglect all contributions $\propto v_I$.
This will be a good approximation for quasi-particles with wavelengths far below the healing length
$\xi=1/\sqrt{8\pi a_s \rho_{\rm c}}$ where $a_s$ is the $s$-wave scattering length. 
In this limit, we get the simple equation 
\bea
\label{eq:master-decay}
\frac{d\hat\varrho_{\chi}}{dt}
=
\sum_I \gamma_{I} \Big( \hat{b}_I \hat\varrho_{\chi} \hat{b}_I^\dagger - 
\frac12 \left\{  \hat{b}_I^\dagger  \hat{b}_I  ,\hat\varrho_{\chi} \right\} \Big)  
\,,
\eea
with the mode-dependent decay rates 
\bea
\gamma_{I} = 9\Gamma N_{\rm c}^2 \int d^3r |\psi_{\rm c}(\mbf{r})|^4 |u_I(\mbf{r})|^2 
\,.
\eea
For quantum enhanced sensing, interesting observables are the quadratures, e.g., 
the positions $\hat X_I=\hat{b}_I^\dagger+\hat{b}_I$
(note that this quantity is often defined with a factor of $1/\sqrt{2}$).
According to Eq.~\eqref{eq:master-decay}, their variances evolve as 
\bea
\label{eq:decay+noise}
\frac{d}{dt}\langle\hat X_I^2\rangle=-\gamma_I\langle\hat X_I^2\rangle+\gamma_I
\,.
\ea
In addition to the usual decay term $-\gamma_I\langle\hat X_I^2\rangle$, 
we get a noise term $+\gamma_I$ stemming from the unavoidable quantum fluctuations 
encoded in the commutator of $\hat{b}_I^\dagger$ and $\hat{b}_I$, as expected from 
the fluctuation-dissipation theorem. 

Now, for sensing applications which can be translated to measuring the generalized position variable 
$\hat X_I$ to high accuracy, a popular scheme of quantum enhanced sensing is to prepare a squeezed state 
with reduced uncertainty $\langle\hat X_I^2\rangle_{\rm{in}}\ll 1$ in that 
direction\footnote{Of course, the uncertainty in the other direction 
$\langle\hat P_I^2\rangle_{\rm{in}}\gg 1$ would be correspondingly larger.} 
as the initial state. 
Then Eq.~\eqref{eq:decay+noise} implies that the variance $\langle\hat X_I^2\rangle$ is doubled after 
a relatively short time $t\approx\langle\hat X_I^2\rangle_{\rm{in}}/\gamma_I$.
In other words, we find a comparably fast deterioration of accuracy, which limits the maximum sensitivity: 
The higher the desired accuracy, the more squeezing is necessary, and thus the faster its deterioration. 

\subsection{Squeezing versus decay}\label{sec:squeezing-decay}

Having found such a comparable fast smearing out of the initially narrow variance
$\langle\hat X_I^2\rangle_{\rm{in}}\ll 1$, one could try to counteract this process by continuously 
squeezing the state in order to reduce $\langle\hat X_I^2\rangle$ or keep it small. 
Thus, let us consider the Hamiltonian 
\bea
\label{eq:Hamiltonian-sqeezing}
\hat H_0^S
= -i\sum_I
\frac{\Xi_I}{2} 
\left[\left(\hat b_I^\dagger\right)^2-\hat b_I^2\right] 
\ea
generating single-mode squeezing for all modes $I$ with the rates $\Xi_I$.
The resulting master equation 
\bea
\label{eq:master-squeezing+decay}
\frac{d\hat\varrho_{\chi}}{dt}
=
-i\left[\hat H_0^S ,\hat\varrho_{\chi}\right] +
\sum_I \gamma_{I} \Big( \hat{b}_I \hat\varrho_{\chi} \hat{b}_I^\dagger - 
\frac12 \left\{  \hat{b}_I^\dagger  \hat{b}_I  ,\hat\varrho_{\chi} \right\} \Big)  
\,,
\nn
\eea
can be solved for each mode $I$ independently. 
The variances can be obtained easily (see also \cite{Schuetzhold_2005,dodonov1998dynamical}) 
\bea
\label{eq:squeezing+decay+noise}
\frac{d}{dt}\langle\hat X_I^2\rangle=-(2\Xi_I+\gamma_I)\langle\hat X_I^2\rangle+\gamma_I
\,.
\ea
We see that maintaining small variances $\langle\hat X_I^2\rangle\ll1$ requires rather 
strong squeezing 
\bea
\Xi_I\approx\frac{\gamma_I}{2\langle\hat X_I^2\rangle}
\,.
\ea
These strong squeezing rates must be realized experimentally by externally driving the 
system, which poses restrictions on the potential of quantum enhanced sensing. 
As another point, such a strong squeezing would decrease one variance $\langle\hat X_I^2\rangle$
but increase the other $\langle\hat P_I^2\rangle$ and thus generate strong excitations.

\subsection{Rotating-wave approximation}\label{sec:rotating-wave} 

Let us now venture a few steps towards a more realistic description of phonon modes, especially those with 
wavelengths larger or comparable to the healing length.  
To this end, we include the undisturbed (Bogoliubov-de~Gennes) Hamiltonian of the phonon modes 
\bea
\label{eq:Hamiltonian-omega}
\hat H_0^R
=\sum_I\omega_I\hat b_I^\dagger\hat b_I
\,,
\ea
with the phonon-mode eigen-frequencies $\omega_I$, and include the previously omitted terms $\propto v_I$.  
Assuming that the phononic frequency scales $\omega_I\pm\omega_J$ are much faster than the 
Lindblad dynamics, we may use the rotating-wave approximation and arrive at 
\bea
\label{eq:masterunif}
\nonumber 
\frac{d\hat\varrho_{\chi}}{dt}
&=& 
-i\left[\hat H_0^R ,\hat\varrho_{\chi}\right] 
+  \sum_I \gamma_{I}^u \Big( \hat{b}_I \hat\varrho_{\chi} \hat{b}_I^\dagger - 
\frac12 \left\{  \hat{b}_I^\dagger  \hat{b}_I  ,\hat\varrho_{\chi} \right\} \Big)  
\\
&& 
+  \sum_I \gamma_{I}^v \Big( \hat{b}_I^\dagger \hat\varrho_{\chi} \hat{b}_I 
- \frac12 \left\{  \hat{b}_I  \hat{b}_I^\dagger  ,\hat\varrho_{\chi} \right\} \Big)  
\,,
\eea
where the cooling and heating terms are 
\bea
\label{eq:cooling}
\gamma_{I}^u &=& 9\Gamma N_{\rm c}^2 \int d^3r |\psi_{\rm c}(\mbf{r})|^4 |u_I(\mbf{r})|^2 
\,,
\\
\gamma_{I}^v &=& 9\Gamma N_{\rm c}^2 \int d^3r |\psi_{\rm c}(\mbf{r})|^4 |v_I(\mbf{r})|^2 
\label{eq:heating}
\,.
\eea
Consistent with the rotating-wave approximation, we consider the co-rotating position quadrature 
of a single mode $\hat{X}_I(t)=\hat b_{I}^\dagger e^{-i\omega_{I} t} + \hat b_{I} e^{i\omega_{I} t}$,  
whose variance evolves as
\bea
\label{eq:dglquad}
\frac{d}{dt}\langle\hat{X}_I^2(t) \rangle 
&=&  
- \gamma_I^- \langle \hat{X}_I^2(t) \rangle  + \gamma_{I}^+
\,,
\eea
where $\gamma_I^\pm = \gamma_{I}^u \pm \gamma_{I}^v$. 
Consequently, the conclusions are qualitatively the same as in the previous scenarios. 
As a main difference, the cooling $\gamma_{I}^u$ and heating $\gamma_{I}^v$ terms both 
contribute equally to the noise $\gamma_{I}^+$ while they act against each other for the 
decay rate $\gamma_I^-$. 

\subsection{Homogeneous condensates}\label{sec:homogeneous} 

In order to obtain more explicit expressions, we have to specify the condensate wave-function. 
In the following, we assume an approximately homogeneous condensate. 
Without loss of generality, we set the total chemical potential to zero such that the condensate 
wave-function becomes (approximately) spatially and temporally constant $\psi_{\rm c}\approx\rm{const}$. 
In this case, the modes $I$ can be labeled by their wave-numbers $\mathbf k$ and the frequencies
are
\bea
\omega_{\mathbf{k}}
=
\frac{|\mathbf{k}|}{2m}\sqrt{\frac{2}{\xi^2}+\mathbf{k}^2}
=
c_{\rm s}|\mathbf{k}|\sqrt{1+\frac12\xi^2\mathbf{k}^2}
\,.
\ea
Again $\xi=1/\sqrt{8\pi a_s \rho_{\rm c}}$ is the healing length, which can also be written as 
$\xi=1/\sqrt{2mg\rho_{\rm c}}$ in terms of the coupling constant $g=4\pi a_s/m$ of the condensate atoms. 
Furthermore, $c_{\rm s}$ denotes the speed of sound given by $1/c_{\rm s}=\sqrt{2}m\xi$ or 
$c_{\rm s}=\sqrt{g\rho_{\rm c}/m}$. 

Apart from the quantization volume normalization, the mode functions $u_\mathbf{k}$ and 
$v_\mathbf{k}$ are given by plane (propagating or standing) waves with the 
Bogoliubov coefficients $\alpha_\mathbf{k}$ and $\beta_\mathbf{k}$ as pre-factors, where 
$\alpha_\mathbf{k}=(\sigma_\mathbf{k}^{-1}+\sigma_\mathbf{k})/2$ and 
$\beta_\mathbf{k}=(\sigma_\mathbf{k}^{-1}-\sigma_\mathbf{k})/2$ with 
\bea
\label{eq:sigma}
\sigma_\mathbf{k}=\sqrt[4]{1+\frac{2}{\xi^2\mathbf{k}^2}}
\,.
\ea
The damping constants \eqref{eq:cooling}-\eqref{eq:heating} reduce to 
$\gamma_\mathbf{k}^u=\alpha_\mbf{k}^2\gamma$ and 
$\gamma_\mathbf{k}^v=\beta_\mbf{k}^2\gamma$, respectively, where $\gamma = 3 D \rho_{\rm c}^2$, 
which implies $\gamma_\mathbf{k}^- =(\alpha_\mbf{k}^2 - \beta_\mbf{k}^2)\gamma = \gamma$ and 
$\gamma_\mathbf{k}^+ =(\alpha_\mbf{k}^2 + \beta_\mbf{k}^2)\gamma\geq\gamma$.  

For large $|\mathbf{k}|\gg 1/\xi$, i.e., in the free-particle limit 
$\omega_{\mathbf{k}}\to\mathbf{k}^2/(2m)$, 
we find $\alpha_\mathbf{k}\to1$ and $\beta_\mathbf{k}\to0$
as indicated above and $\gamma_\mathbf{k}^\pm\to\gamma$. 
Instead for small $|\mathbf{k}|\ll 1/\xi$, i.e., in the phonon limit 
$\omega_{\mathbf{k}}\to c_{\rm s}|\mathbf{k}|$, both $\alpha_\mbf{k}^2$ and 
$\beta_\mbf{k}^2$ grow as $\sqrt{2}/(4\xi|\mbf{k}|)$.
This leads to $\gamma_\mathbf{k}^+ \to \gamma (\sqrt{2}\xi|\mbf{k}|)^{-1}$,
and the noise term is significantly amplified.

\subsection{Squeezed reference frame}\label{sec:squeezed} 

Actually, one may understand the underlying dynamics even without invoking the rotating-wave 
approximation employed in Sec.~\ref{sec:rotating-wave}. 
Considering a homogeneous condensate, we may use the Fourier expansion of $\hat\chi$ as our 
mode decomposition in equation (\ref{Bogoliubov-trafo}). 
Inserting this into the master equation~\eqref{eq:masterpho0}, we get 
\bea
\label{eq:master-Fourier}
\frac{d\hat\varrho_{\chi}}{dt}
= 
-i\left[\hat H_0 ,\hat\varrho_{\chi}\right] 
+  \gamma \sum_\mathbf{k} \Big( \hat{a}_\mathbf{k} \hat\varrho_{\chi} \hat{a}_\mathbf{k}^\dagger 
- \frac12 \left\{  \hat{a}_\mathbf{k}^\dagger  \hat{a}_\mathbf{k} ,\hat\varrho_{\chi} \right\} \Big)  
\,,
\nn
\eea
in terms of the original atomic creation and annihilation operators $\hat{a}_\mathbf{k}^\dagger$
and $\hat{a}_\mathbf{k}$.
For homogeneous condensates, we obtain a $\mathbf{k}$-independent decay rate $\gamma$ 
from Eq.~\eqref{eq:masterpho0}.

The atomic operators $\hat{a}_\mathbf{k}^\dagger$ and $\hat{a}_\mathbf{k}$ 
are related to the phononic quasi-particle operators $\hat{b}_\mathbf{k}^\dagger$
and $\hat{b}_\mathbf{k}$ via the Bogoliubov transformation~\eqref{Bogoliubov-trafo}
which corresponds to a unitary squeezing operation $\hat U_\mathbf{k}$ for each 
mode\footnote{Using the mode functions $\exp\{i\mathbf{k}\cdot\mathbf{r}\}$
for periodic boundary conditions, Eq.~\eqref{unitary} would read 
$\hat{a}_\mathbf{k}
=
\alpha_\mathbf{k}\hat{b}_\mathbf{k}+\beta_\mathbf{k}\hat{b}_{-\mathbf{k}}^\dagger$. 
For reflecting boundary conditions (e.g., Dirichlet or Neumann), one would use sine or 
cosine functions instead where $\mathbf{k}$ and $-\mathbf{k}$ correspond to the same mode
and thus we may write 
$\hat{a}_\mathbf{k}
=
\alpha_\mathbf{k}\hat{b}_\mathbf{k}+\beta_\mathbf{k}\hat{b}_{\mathbf{k}}^\dagger$
as in Eq.~\eqref{unitary}.} 
\bea
\label{unitary}
\hat{a}_\mathbf{k}
=
\alpha_\mathbf{k}\hat{b}_\mathbf{k}+\beta_\mathbf{k}\hat{b}_\mathbf{k}^\dagger
=
\hat U_\mathbf{k}^\dagger\hat{b}_\mathbf{k}\hat U_\mathbf{k}
\,.
\ea
If we directly insert the above Bogoliubov transformation 
$\hat{a}_\mathbf{k}=\alpha_\mathbf{k}\hat{b}_\mathbf{k}+\beta_\mathbf{k}\hat{b}_\mathbf{k}^\dagger$
into Eq.~\eqref{eq:master-Fourier} and neglect all counter-rotating terms such as $\hat{b}_\mathbf{k}^2$,
where the time-evolution is generated by Eq.~\eqref{eq:Hamiltonian-omega}, we recover 
Eq.~\eqref{eq:masterunif} for homogeneous condensates. 
In this form, the $\mathbf k$-dependence of the Bogoliubov coefficients 
$\alpha_\mathbf{k}$ and $\beta_\mathbf{k}$ entails a $\mathbf k$-dependence of the rates 
$\gamma_\mathbf{k}^u$ and $\gamma_\mathbf{k}^v$ discussed after Eq.~\eqref{eq:sigma}.

However, another representation can be more convenient:
Since the original atomic operators $\hat{a}_\mathbf{k}$ are basically the Lindblad jump operators in 
Eq.~\eqref{eq:master-Fourier}, we can use them as a basis and express everything -- 
including the Hamiltonian $\hat H_0$ -- in terms of 
$\hat{a}_\mathbf{k}^\dagger$ and $\hat{a}_\mathbf{k}$ instead of the phononic operators 
$\hat{b}_\mathbf{k}^\dagger$ and $\hat{b}_\mathbf{k}$. 
Assuming that $\hat H_0$ is a bi-linear function of the $\hat{a}_\mathbf{k}^\dagger$ and 
$\hat{a}_\mathbf{k}$ (or, equivalently, of the $\hat{b}_\mathbf{k}^\dagger$ and $\hat{b}_\mathbf{k}$)
which does not couple the different modes $\mathbf{k}$ with each other, its most general form reads 
\bea
\label{eq:squeezed-Hamiltonian}
\hat H_0
=
\sum_\mathbf{k}
\left(
\tilde\omega_\mathbf{k}\hat{a}_\mathbf{k}^\dagger\hat{a}_\mathbf{k}
-\frac{i}{2}\,\tilde\Xi_\mathbf{k}(\hat{a}_\mathbf{k}^\dagger)^2
+\frac{i}{2}\,\tilde\Xi_\mathbf{k}^*\hat{a}_\mathbf{k}^2
\right) 
\,.
\ea
Note that, even if the Hamiltonian $\hat H_0$ was diagonal in terms of the phononic operators 
$\hat{b}_\mathbf{k}^\dagger$ and $\hat{b}_\mathbf{k}$, i.e., had the form of $\hat H_0^R$ 
in Eq.~\eqref{eq:Hamiltonian-omega}, it would still contain non-zero $\tilde\Xi_\mathbf{k}$-terms
in the representation~\eqref{eq:squeezed-Hamiltonian}. 
One way to obtain them would be to insert the inverse Bogoliubov transformation~\eqref{unitary},
i.e., $\hat{b}_\mathbf{k}=\hat U_\mathbf{k}\hat{a}_\mathbf{k}\hat U_\mathbf{k}^\dagger$,
into the form $\hat{b}_\mathbf{k}^\dagger\hat{b}_\mathbf{k}$. 
However, as the Bogoliubov transformation~\eqref{unitary} is precisely chosen in order to diagonalize 
the original Bogoliubov-de~Gennes Hamiltonian $\hat H_0$, we can find those $\tilde\Xi_\mathbf{k}$-terms 
directly in $\hat H_0$ when it is expressed in terms of the atomic operators: 
Inserting the mean-field split~\eqref{eq:split} into the atomic Hamiltonian $\hat H_0$ containing the 
interaction term $(\hat\Psi^\dagger)^2\hat\Psi^2$, we also get $\hat\chi^2$ and $(\hat\chi^\dagger)^2$
terms in the effective Bogoliubov-de~Gennes Hamiltonian 
which translate into $\hat{a}_\mathbf{k}^2$ and $(\hat{a}_\mathbf{k}^\dagger)^2$ contributions after 
a Fourier transform. 
In this case, we would get constant $\tilde\Xi_\mathbf{k}=\tilde\Xi$ for a homogeneous condensate. 

In order to study the dynamics following from Eqs.~\eqref{eq:master-Fourier} and 
\eqref{eq:squeezed-Hamiltonian}, we consider the expectation values of the 
number operator $\langle\hat{a}_\mathbf{k}^\dagger\hat{a}_\mathbf{k}\rangle$  
and the anomalous average 
$\langle\hat{a}_\mathbf{k}^2\rangle$
as well as its complex conjugate. 
Their evolution equations read 
\bea
\frac{d}{dt}\langle\hat{a}_\mathbf{k}^\dagger\hat{a}_\mathbf{k}\rangle
=
-\tilde\Xi_\mathbf{k}\langle(\hat{a}_\mathbf{k}^\dagger)^2\rangle - 
\tilde\Xi_\mathbf{k}^*\langle\hat{a}_\mathbf{k}^2\rangle 
-\gamma\langle\hat{a}_\mathbf{k}^\dagger\hat{a}_\mathbf{k}\rangle
\,,
\ea
and for the anomalous average
\bea
\frac{d}{dt}\langle\hat{a}_\mathbf{k}^2\rangle
=
-2i\tilde\omega_\mathbf{k}\langle\hat{a}_\mathbf{k}^2\rangle
-2\tilde\Xi_\mathbf{k}(\langle\hat{a}_\mathbf{k}^\dagger\hat{a}_\mathbf{k}\rangle+1) 
-\gamma\langle\hat{a}_\mathbf{k}^2\rangle
\,.
\ea
Introducing the vector 
$\mathbf{w}=(
\langle\hat{a}_\mathbf{k}^2\rangle,
\langle\hat{a}_\mathbf{k}^\dagger\hat{a}_\mathbf{k}\rangle,
\langle\hat{a}_\mathbf{k}^2\rangle^*)$, 
this linear system can be cast into the form 
$\dot{\mathbf{w}}+\mathbf{M}\cdot\mathbf{w}=\mathbf{s}$ 
with the source 
$\mathbf{s}=-2(\tilde\Xi_\mathbf{k},0,\tilde\Xi_\mathbf{k}^*)$
and the $3\times3$-matrix
\bea
\mathbf{M}=
\left(
\begin{array}{ccc}
{\gamma} +2i\tilde\omega_\mathbf{k} & 2\tilde\Xi_\mathbf{k} & 0 
\\
\tilde\Xi_\mathbf{k}^* & {\gamma} & \tilde\Xi_\mathbf{k}
\\
0 & 2\tilde\Xi_\mathbf{k}^* & {\gamma} -2i\tilde\omega_\mathbf{k}
\end{array}
\right)
\,,
\ea
with the eigenvalues ${\gamma}$ and 
${\gamma}\pm2\sqrt{|\tilde\Xi_\mathbf{k}|^2-\tilde\omega_\mathbf{k}^2}$. 
Thus, this set of equations can be solved explicitly
-- incorporating the results of the previous sections as limiting cases. 
For example, for $\tilde\Xi_\mathbf{k}=\tilde\omega_\mathbf{k}=\beta_\mathbf{k}=0$,
we recover Eq.~\eqref{eq:decay+noise}.
Note that the above matrix equation just yields the exponential decay 
$\propto\exp\{-{\gamma} t\}$ of the expectation values 
$\langle\hat{a}_\mathbf{k}^\dagger\hat{a}_\mathbf{k}\rangle$, 
$\langle\hat{a}_\mathbf{k}^2\rangle$, and $\langle\hat{a}_\mathbf{k}^2\rangle^*$,
the noise term in Eq.~\eqref{eq:decay+noise} stems from the commutator of the operators 
$\hat{a}_\mathbf{k}$ and $\hat{a}_\mathbf{k}^\dagger$ 
when re-expressing the variance
$\langle\hat{X}_\mathbf{k}^2\rangle$ as a function of those expectation values 
$\langle\hat{a}_\mathbf{k}^\dagger\hat{a}_\mathbf{k}\rangle$, 
$\langle\hat{a}_\mathbf{k}^2\rangle$, and $\langle\hat{a}_\mathbf{k}^2\rangle^*$. 
This commutator represents the quantum fluctuations:
While the expectation values 
$\langle\hat{a}_\mathbf{k}^\dagger\hat{a}_\mathbf{k}\rangle$, 
$\langle\hat{a}_\mathbf{k}^2\rangle$, and $\langle\hat{a}_\mathbf{k}^2\rangle^*$
tend to zero under the influence of the decay channel, this is not true for the 
variance $\langle\hat{X}_\mathbf{k}^2\rangle$, which tends to unity -- 
respecting the Heisenberg uncertainty principle.

In analogy, one may derive a closed set of equations for the linear expectation values 
$\langle\hat{a}_\mathbf{k}^\dagger\rangle$ and $\langle\hat{a}_\mathbf{k}\rangle$. 
Apart from the $\hat H_0$-evolution, 
we obtain additional decay terms 
$-\gamma\langle\hat{a}_\mathbf{k}^\dagger\rangle/2$ and 
$-\gamma\langle\hat{a}_\mathbf{k}\rangle/2$. 
Note that we did not assume Gaussian states so far --  this will be the subject of the 
next section.

\section{Gaussian States}\label{sec:gaussian}

In the following, we restrict our considerations to Gaussian states $\hat\varrho_\chi$.
To this end, we assume that the undisturbed quasi-particle Hamiltonian $\hat H_0$ in 
Eq.~\eqref{eq:masterpho0} is -- at least to a sufficiently good approximation -- given by a linear 
combination of terms linear or bi-linear in the operators $\hat{b}_I$ and $\hat{b}_I^\dagger$. 
Combining these creation and annihilation operators in an operator-valued pseudo-vector $\hat\zeta_I$, 
the general structure of $\hat H_0^{\rm{Gauss}}$ is thus 
\bea
\label{eq:Hamiltonian(bi)linear}
\hat H_0^{\rm{Gauss}}
=\sum_{IJ}\hat\zeta_I\Omega_{IJ}\hat\zeta_J+\sum_I\hat\zeta_I\delta_I
\,,
\ea
where the block matrix $\Omega_{IJ}$ contains the single-mode $\Xi_I$ and multi-mode $\Xi_{IJ}$ 
squeezing rates as well as the single-mode frequencies $\omega_I$ and possible multi-mode rotations 
$\omega_{IJ}$, while the $\delta_I$ terms generate coherent displacements. 
Note that the transformations $\hat U_\mathbf{k}$ to the squeezed reference frame mentioned in the 
previous Section preserve this structure and just change the parameters $\Omega_{IJ}\to\tilde\Omega_{IJ}$
and $\delta_I\to\tilde\delta_I$.

Under that condition~\eqref{eq:Hamiltonian(bi)linear}, the Lindblad master equation~\eqref{eq:masterpho0} 
preserves Gaussianity, i.e., it maps initial Gaussian states $\hat\varrho_\chi(t=0)$ to final Gaussian 
states $\hat\varrho_\chi(t)$, see also \cite{Serafini_2005}. 
Gaussian states strongly simplify the analysis. 
They are uniquely defined by their displacement vector 
\bea
D_I=\langle\hat\zeta_I\rangle 
\ea
and the covariance matrix
\bea
\Sigma_{IJ}=
\langle\hat\zeta_I\hat\zeta_J^\dagger+\hat\zeta_J^\dagger\hat\zeta_I\rangle-2D_ID_J^* 
\,,
\ea
which are the generalization of the expectation values 
$\langle\hat{b}_\mathbf{k}^\dagger\rangle$ and $\langle\hat{b}_\mathbf{k}\rangle$
as well as 
$\langle\hat{b}_\mathbf{k}^\dagger\hat{b}_\mathbf{k}\rangle$, 
$\langle\hat{b}_\mathbf{k}^2\rangle$ and $\langle\hat{b}_\mathbf{k}^2\rangle^*$
(or, equivalently, 
$\langle\hat{a}_\mathbf{k}^\dagger\rangle$ and $\langle\hat{a}_\mathbf{k}\rangle$
as well as 
$\langle\hat{a}_\mathbf{k}^\dagger\hat{a}_\mathbf{k}\rangle$, 
$\langle\hat{a}_\mathbf{k}^2\rangle$ and $\langle\hat{a}_\mathbf{k}^2\rangle^*$)
discussed in the previous Section. 
All higher-order expectation values can be reduced to these two quantities 
(i.e., the first and second cumulants) via a Wick type expansion. 

\subsection{Parameter estimation}\label{sec:parameter}

The simple structure of Gaussian states discussed above facilitates several calculations which can be 
extremely hard for general states. 
For instance, because general mixed quantum states can be decomposed into pure states in many different ways, 
it is often necessary to minimize over all possible decompositions (which is referred to as convex roof 
construction).  
One example are entanglement measures, another example is the Quantum Fisher Information, 
see, e.g., \cite{helstrom1976quantum,holevo1982probabilistic,Braunstein:1994}. 
This quantity measures how much a state changes if we modify an external parameter $\vartheta$ which 
occurs in the Hamiltonian $\hat H_0$, for example. 
Of course, this is relevant for estimating this external parameter $\vartheta$ by measuring the state.

In terms of the $\vartheta$-dependent displacement vector $\mathbf{D}_\vartheta$ and covariance matrix 
$\mathbf{\Sigma}_\vartheta$ at a given time $t$, the Quantum Fisher Information 
(for a Gaussian state) reads \cite{pinel2013quantum}
\bea
\label{eq:QFI}
F_\vartheta
&=&
\frac12\,\frac{\Tr\{(\mathbf{\Sigma}_\vartheta^{-1}\cdot\mathbf{\Sigma}_\vartheta')^2\}}{1+P_\vartheta^2}
+2\frac{(P_\vartheta')^2}{1-P_\vartheta^4}+
\nn
&&
+2(\mathbf{D}_\vartheta')^*\cdot\mathbf{\Sigma}_\vartheta^{-1}\cdot\mathbf{D}_\vartheta'
\,.
\ea
Here $P_\vartheta=1/\sqrt{\det\mathbf{\Sigma}_\vartheta}$ denotes the purity of the state and primes 
denote derivatives with respect to $\vartheta$. 
These derivatives describe how much the state changes when modifying the external parameter $\vartheta$ 
and the Quantum Fisher Information~\eqref{eq:QFI}
quantifies the potential to estimate this change by measuring the state.
The achievable accuracy is then given by the Quantum Cram\'er-Rao bound 
\cite{helstrom1976quantum,holevo1982probabilistic,Braunstein:1994} 
on the minimum variance of $\vartheta$
\bea
(\Delta\vartheta)^2\geq\frac{1}{F_\vartheta}\,,
\ea
where we assumed a single measurement. 
Thus, for high accuracy, a large Quantum Fisher Information~\eqref{eq:QFI} is necessary. 
One way to achieve this would be to increase the derivatives $\mathbf{\Sigma}_\vartheta'$, 
$P_\vartheta'$, and $\mathbf{D}_\vartheta'$. 
However, this is limited by the experimentally available coupling strengths etc.
The idea of quantum enhanced sensing\footnote{In principle, one could also try to estimate $\vartheta$ 
via its impact on the purity $P_\vartheta'$, e.g., by having the constants $\gamma_I$ in the 
Lindblad master equation depend on $\vartheta$. 
However, preparing such a coupling and estimating $\vartheta$ in that way seems extremely challenging 
and thus we do not discuss this option here.
The potential for enhanced sensing could then be identified with the $1-P_\vartheta^4$ 
in the denominator in~\eqref{eq:QFI}, which implies that the fraction could be very large for 
purities $P_\vartheta$ near unity (i.e., almost pure states). 
Since decoherence tends to decrease the purity, it would also limit the enhancement potential for 
this scheme.}
is to modify the covariance matrix $\mathbf{\Sigma}_\vartheta$
such that one or some of its eigenvalues are small and thus changes $\mathbf{\Sigma}_\vartheta'$ or 
$\mathbf{D}_\vartheta'$ in that direction (i.e., the corresponding eigenvectors) are enhanced by 
the multiplication with $\mathbf{\Sigma}_\vartheta^{-1}$. 

\subsection{Decoherence}\label{sec:gauss-decoherence}

As one would already expect from the results of Sec.~\ref{sec:decoherence}, the impact of decoherence tends 
to limit the potential of quantum enhanced sensing. 
In the framework discussed above, this manifests itself in the growth of the small eigenvalues of the 
covariance matrix $\mathbf{\Sigma}_\vartheta$. 
If we consider the pure decay channel in Sec.~\ref{sec:decay} or the evolution within the squeezed 
reference frame in Sec.~\ref{sec:squeezed}, 
we find the equation of motion for the covariance matrix
\bea
\label{eq:eom-covariance-matrix}
\frac{d}{dt}\mathbf{\Sigma}_\vartheta
&=&
\mathbf{\Omega}\cdot\mathbf{\Sigma}_\vartheta+\mathbf{\Sigma}_\vartheta\cdot\mathbf{\Omega}^\dagger
\nn
&&
-\frac12
\left(
\mathbf{\Gamma}\cdot\mathbf{\Sigma}_\vartheta+\mathbf{\Sigma}_\vartheta\cdot\mathbf{\Gamma}
\right)
+\mathbf{\Gamma}
\,.
\ea
Here $\mathbf{\Omega}$ represents the Hamiltonian evolution and is built from the rates $\Omega_{IJ}$ 
in Eq.~\eqref{eq:Hamiltonian(bi)linear}, while $\mathbf{\Gamma}$ is a diagonal matrix containing 
all the decay rates $\gamma_I$. 
Thus, if $\mathbf{u}_\lambda$ is an eigenvector of $\mathbf{\Sigma}_\vartheta$ with a small eigenvalue
$\lambda$, we find that the rate of change of the projection of $\mathbf{\Sigma}_\vartheta$ onto that 
eigenvector $\mathbf{u}_\lambda\cdot\mathbf{\Sigma}_\vartheta\cdot\mathbf{u}_\lambda$ scales with the 
small eigenvalue $\lambda$ plus the noise term 
$\mathbf{u}_\lambda\cdot\mathbf{\Gamma}\cdot\mathbf{u}_\lambda$.
Since $\mathbf{\Gamma}$ is a positive matrix (assuming that all $\gamma_I$ are positive), we see that 
this poses a quite general limitation of the achievable accuracy. 

For the rotating-wave approximation discussed in Sec.~\ref{sec:rotating-wave}, we find a very similar 
evolution equation 
\bea
\label{eq:oem-rotating-wave}
\frac{d}{dt}\mathbf{\Sigma}_\vartheta
=
-\frac12
\left(
\mathbf{\Gamma}_-\cdot\mathbf{\Sigma}_\vartheta+\mathbf{\Sigma}_\vartheta\cdot\mathbf{\Gamma}_-
\right)
+\mathbf{\Gamma}_+
\,,
\ea
where $\mathbf{\Gamma}_\pm$ are again diagonal matrices containing the decay rates $\gamma_I^\pm$ 
in complete analogy to Eq.~\eqref{eq:dglquad}. 
Note that we did not include the Hamiltonian evolution $\mathbf{\Omega}$ since the above equation 
refers to the rotating reference frame. 
As another difference to the evolution equation~\eqref{eq:eom-covariance-matrix}, the decoherence 
does not drive the system towards the ground state (where $\mathbf{\Sigma}_\vartheta$ would be given 
by the identity matrix), but to the asymptotic state 
$\mathbf{\Sigma}_\infty=\mathbf{\Gamma}_+\cdot\mathbf{\Gamma}_-^{-1}$ corresponding to detailed balance,
see equation~(\ref{eq:Sigmat}) in the Appendix for the solution of~(\ref{eq:oem-rotating-wave}).
Apart from that, the main conclusions remain unchanged. 

\section{single-mode case}\label{sec:single-mode}

For a single mode, the most general pure Gaussian state is a squeezed, displaced and rotated state,
which can be obtained from the ground state $\ket{0}$ via the unitary operations 
$\hat S$, $\hat D$, and $\hat R$ generated by the respective contributions to the Hamiltonian 
$\hat H_S=-i\Xi(\hat b^\dagger)^2/2+\rm{h.c.}$, $\hat H_D=i\delta\hat b^\dagger + \rm{h.c.}$, 
and $\hat H_R=\omega\hat b^\dagger\hat b$. 
Their symplectic representations are $2\times2$-matrices acting on the covariance matrix 
$\mathbf{\Sigma}$ and displacement vector $\mathbf{D}$, such as 
$\mathbf{\Sigma}\to\mathbf{S}(r)\cdot\mathbf{\Sigma}\cdot\mathbf{S}^\dagger(r)$  
with the squeezing matrix 
\bea
\mbf{S}(r) = 
\left( 
\begin{array}{cc}  
\cosh(r)   & -\sinh(r) 
\\
-\sinh(r)   &  \cosh(r) 
\end{array} 
\right)
\,,
\eea
for the case of real $\Xi=r\in\mathbb R$.
Taking the ground state $\ket{0}$ with $\mathbf{\Sigma}=\mathbf{1}$ as the initial state,  
one finds 
\bea
\label{eq:initial-squeezed}
\mbf{\Sigma}(r) = 
\left( 
\begin{array}{cc}  
\cosh(2r)   & -\sinh(2r) 
\\
-\sinh(2r)   &  \cosh(2r) 
\end{array} 
\right)
\,,
\eea
with the two eigenvalues $\lambda_\pm=e^{\pm2r}$. 
Thus, for large squeezing, say  $e^{2r}\gg1$, one can enhance the sensitivity. 

However, as explained above, this pure squeezed state is very vulnerable to decoherence,
which adds noise and thereby turns it into a mixed state. 
For example, the evolution equation~\eqref{eq:oem-rotating-wave} with the above 
state~\eqref{eq:initial-squeezed} as the initial state $\mbf{\Sigma}_0$ can be solved as 
\bea
\mbf{\Sigma}(t)=e^{-\gamma^- t}\mbf{\Sigma}_0+\frac{ \gamma^+}{\gamma^-}(1-e^{- \gamma^-t})\bf{1}
\,.
\ea
For short times $t\ll1/ \gamma^-$, we see that the small eigenvalue $\lambda_-$ grows as
\bea
\lambda_-(t)=e^{-2r}+ \gamma^+t-e^{-2r}\gamma^- t+\ord(t^2)
\,.
\ea
Since $e^{-2r}\ll1$ and $ \gamma^+\geq \gamma^-$, the growth term $ \gamma^+t$ dominates and the 
quantum enhanced sensitivity deteriorates after a short time $t=\ord(e^{-2r}/ \gamma^+)$. 
This goes along with a decay of the purity $P(t) \approx  ( 1 + e^{2r} \gamma^+ t )^{-1/2} $ on the same timescale.

\subsection{Displacement Scheme}\label{sec:displacement-scheme}

\begin{figure}
\includegraphics[width=.45\textwidth]{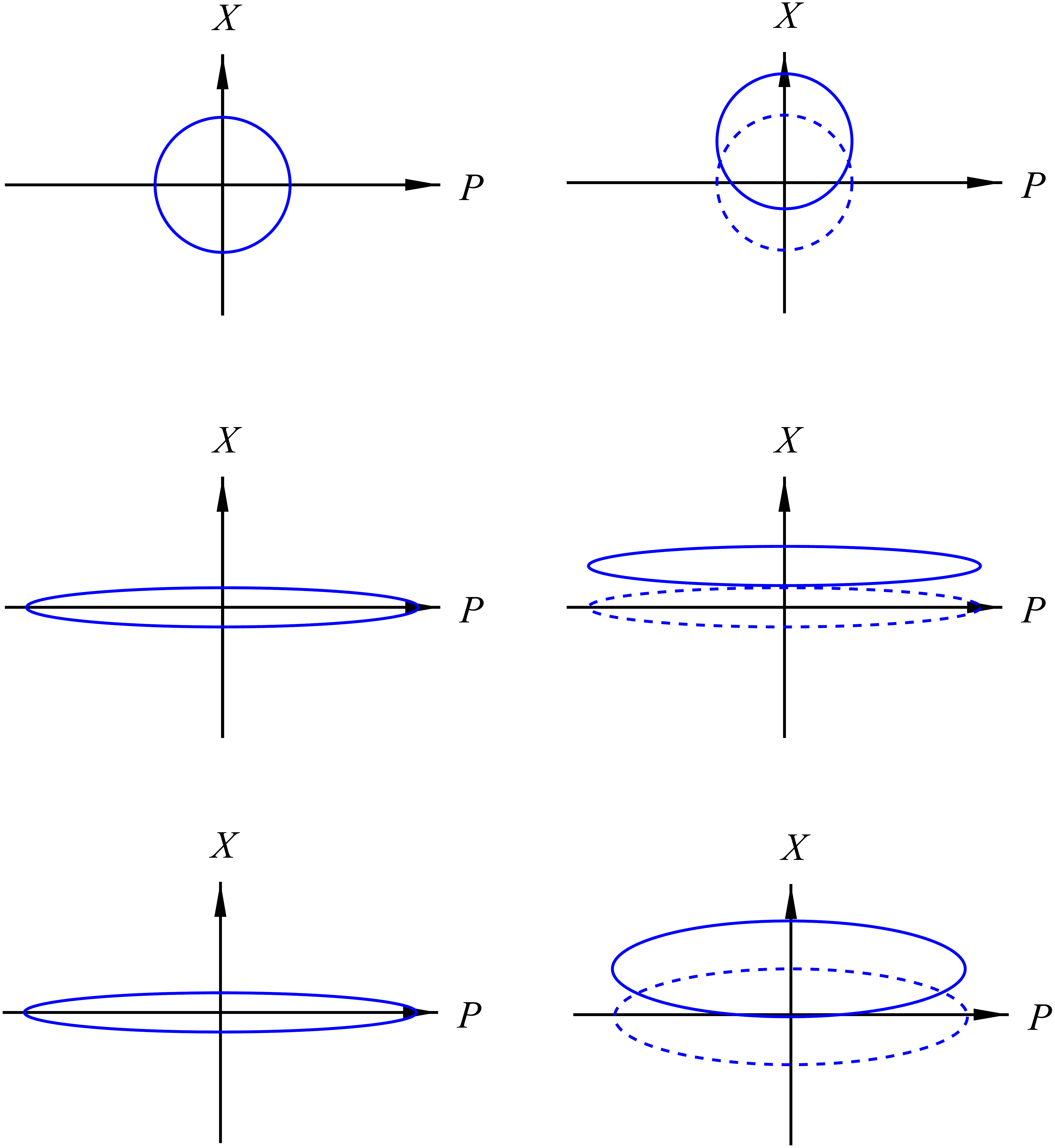} 
\caption{
Sketch (not to scale) of the main idea of quantum enhanced sensing 
(with single-mode squeezing) in the displacement scheme 
and its deterioration due to decoherence. 
Depicted are the Wigner representations of the initial ground state (top left) 
and the coherent state (top right) displaced in $X$-direction $X\to X+\delta$
due to the interaction representing the measurement. 
However, if this displacement $\delta$ is smaller than the initial quantum 
uncertainty $\Delta X$ (radius of the blue circles), the coherent state 
(full blue circle in top right diagram) cannot be distinguished unambiguously
from the initial ground state, i.e., the measurement is not conclusive. 
To overcome this problem, one can start with an initial squeezed state 
(middle left) with a reduced uncertainty $\Delta X$ such that the 
displacement $\delta$ is larger than $\Delta X$ (middle right). 
In the presence of decoherence, however, this reduced initial uncertainty 
$\Delta X$ grows between initialization (bottom left) and final read-out 
(bottom right) such that the state after the interaction (full blue ellipse 
in bottom right diagram) can again not be distinguished unambiguously from 
the state without this interaction (dashed blue ellipse in bottom right diagram).}
\label{figure}
\end{figure}

In order to be more explicit, we have to specify how the signal to be measured is coupled to the 
single mode under consideration.
Encoding the parameter $\vartheta$ to be estimated in a pure displacement $\mathbf{D}_\vartheta$ 
into the direction $\mathbf{u}_{\lambda_-}$ where the variance is small, 
the Quantum Fisher Information~\eqref{eq:QFI} 
simplifies to $F_\vartheta=2|\mathbf{D}_\vartheta'|^2\lambda_-^{-1}$ 
(see Appendix \ref{sec:estdispl} for details). 
Thus, for the initial pure state, the accuracy $\Delta\vartheta$ would be enhanced by the squeezing 
factor $e^{-r}$. 
Adding noise, however, the uncertainty rapidly grows as 
\bea
\label{eq:displacement-decay}
\Delta\vartheta\geq\frac{1}{\sqrt{F_\vartheta}}
=
\frac{e^{-r}}{\sqrt{2}|\mathbf{D}_\vartheta'|}
\to
\frac{e^{-r}}{\sqrt{2}|\mathbf{D}_\vartheta'|}\sqrt{1+e^{2r} \gamma^+t}
\,.
\ea
Let us illustrate the main mechanism by means of a simple example depicted in 
Fig.~\ref{figure}.
The displacement Hamiltonian $\hat H_D$ does not affect the covariance matrix 
$\mathbf{\Sigma}$ at all, but just creates a displacement vector $\mathbf{D}$. 
The impact of decoherence, on the other hand, strongly affects the covariance matrix 
$\mathbf{\Sigma}$, mainly by increasing the small eigenvalue(s), while its influence on the 
displacement vector $\mathbf{D}$ is very weak.
Thus, we may consider the two phenomena quite independently and arrive at the sequence 
shown in Fig.~\ref{figure}.
Here, the small $\lambda_-$ and large $\lambda_+$ eigenvalues of the covariance matrix 
$\mathbf{\Sigma}$ are represented by the minor and major semi-axes of the ellipses 
in the Wigner representation. 
Thus, their area is a measure of the purity (corresponding to the determinant 
of $\mathbf{\Sigma}$).

\subsection{Rotation Scheme}\label{sec:rotation-scheme}

By inspecting Eq.~\eqref{eq:QFI}, we see that the impact of the small eigenvalue $\lambda_-$ 
employed for quantum enhanced sensing is even larger for the first term in Eq.~\eqref{eq:QFI}, 
which scales quadratically in $\lambda_-^{-1}$.
To exploit this scaling, one could encode the parameter $\vartheta$ in the covariance matrix, 
which can be achieved by a rotation $\hat R$, for example. 
The corresponding mechanism is depicted in Fig.~\ref{figure-rot}.
For $e^{2r}\gg1$, 
we obtain (see Appendix \ref{sec:estrot})
\bea
\label{eq:Heisenberg-decay}
\Delta\vartheta\geq\frac{1}{\sqrt{F_\vartheta}}
\approx 
e^{-2r} \sqrt{ 2 + e^{2r}\gamma^+t }
\,.
\ea
%

\begin{figure}
\includegraphics[width=.45\textwidth]{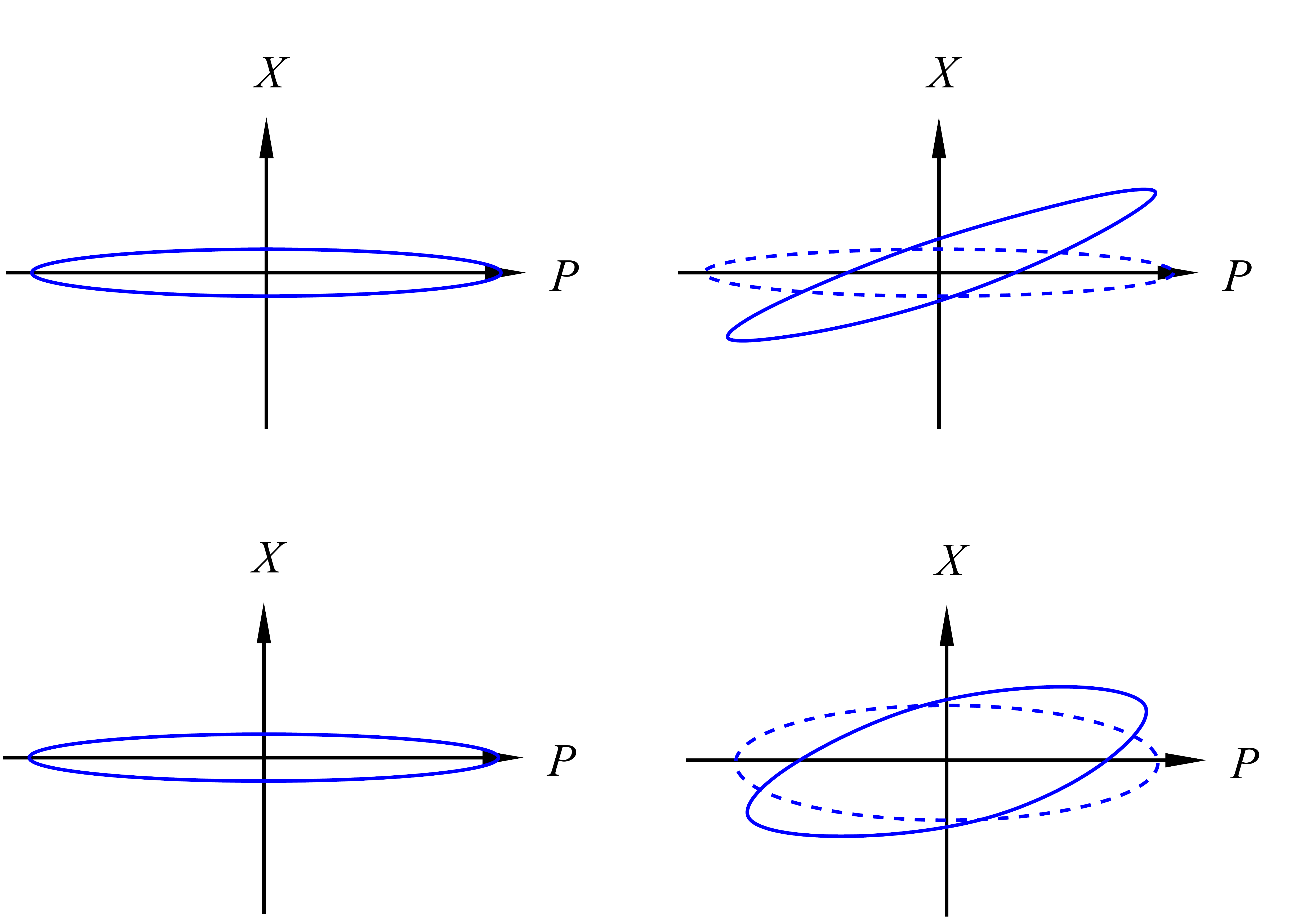} 
\caption{Sketch (not to scale) of the main idea of quantum enhanced sensing 
(with single-mode squeezing) in the rotation scheme 
and its deterioration due to decoherence in analogy to Fig.~\ref{figure}.
The left panels depict the Wigner representations of the initial squeezed states. 
Without decoherence, the rotated squeezed state (solid blue ellipse in top right panel)
has little overlap with the state in the absence of rotation (dashed blue ellipse in top right panel). 
With decoherence, however, their overlap becomes much larger (bottom right panel) 
and thus it is much harder to distinguish them. 
In the squeezing scheme (Sec.~\ref{sec:squeezing-scheme}), we get a qualitatively similar picture. 
}
\label{figure-rot}
\end{figure}

\subsection{Squeezing Scheme}\label{sec:squeezing-scheme}

Finally, one can encode the signal $\vartheta$ also in a squeezing operation itself. 
If we again start from a state squeezed in $X$ direction, as in Figs.~\ref{figure} 
and~\ref{figure-rot}, these figures already suggest that encoding the signal $\vartheta$ 
into a squeezing in exactly the same $X$ direction (or in the orthogonal $P$ direction) 
is not the best idea for achieving quantum enhanced sensing. 
Instead, one should apply a squeezing in a slanting direction (e.g., at $\pm45^\circ$).
Then, for large initial squeezing $e^r\gg1$ and small signal squeezing (i.e., small $\vartheta$), 
we get qualitatively the same picture as in Fig.~\ref{figure-rot}  
(a detailed calculation can be found in Appendix \ref{sec:estsqueezing}). 

\subsection{Sensing continuously encoded parameters}\label{sec:continous}

Above, we considered the simplified situation where the parameter $\vartheta$ is encoded in the 
initial state of the phonon field before its open dynamics leads to deterioration of quantum sensitivity. 
This corresponds to the case of fast encoding of $\vartheta$ in comparison to the time scales given by 
$1/\gamma^-$ and $1/\gamma^+$.
If the time scale of the free dynamics and the duration of the encoding process are similar, 
we have to consider the case of continuous encoding of $\vartheta$, in principle. 
However, for small values of $\vartheta$ and $\gamma_I^- t$, the dynamics of the phonon field differs 
from the case of instantaneous encoding by terms proportional to products and higher powers of $\vartheta$ 
and $\gamma_I^- t$.
Neglecting these higher order contributions, we recover the above results
\footnote{For the case of continuously encoded squeezing, detailed calculations can be found
in Appendix \ref{sec:estcontsqueezing}.}. 
If the rate of change of the encoded parameter $\vartheta$ is to be estimated, 
there exists an optimal measurement duration depending on $\gamma^+$.

\subsection{Heisenberg Limit}\label{sec:Heisenberg-limit}

Note that, for large squeezing $e^{2r}\gg1$, the small eigenvalue $\lambda_-=e^{-2r}$ of the pure state 
scales with the inverse of the number $\langle\hat n\rangle$ of excitation quanta (phonons), 
which is given by $\langle\hat n\rangle=\sinh^2(r)$. 
Thus, the accuracy $\Delta\vartheta$ scales inversely proportional to the number of phonons, which is 
usually referred to as the Heisenberg limit -- in contrast to the usual Poisson limit 
$\Delta\vartheta\sim1/\sqrt{\langle\hat n\rangle}$, which is also referred to as shot-noise limit or standard quantum limit.
However, the deterioration of the Heisenberg limit is not caused by the decay of the excitation quanta -- 
which occurs on a rather long time scale $t=\ord(1/ \gamma^-)$. 
Instead, it is caused by adding noise, which happens on a time scale $t=\ord(e^{-2r}/ \gamma^+)$,
i.e., much faster (for large squeezing).

Note that the Heisenberg scaling $\Delta\vartheta\sim1/\langle\hat n\rangle$ can also be realized via the 
displacement scheme~\eqref{eq:displacement-decay} if we do not just use a strong squeezing to reduce 
$\lambda_-=e^{-2r}$, but also employ a large displacement. 
Of course, a large displacement does also imply a large number of phonons 
$|\mathbf{D}_\vartheta|^2\sim\langle\hat n\rangle$ such that the available signal 
$|\mathbf{D}_\vartheta'|$ scales with $\sqrt{\langle\hat n\rangle}$, giving the total scaling 
$\Delta\vartheta\sim1/\langle\hat n\rangle$.

In both cases, the minimum uncertainty $\Delta\vartheta$ is determined by the Heisenberg limit 
$\Delta\vartheta\sim1/\langle\hat n\rangle$, as expected. 
Note that the number of phonons $\langle\hat n\rangle$ must be much smaller than the total number 
$N_{\rm c}$ of atoms in the condensate for the linearized quasi-particle description to apply.

\section{Experimental Parameters}\label{sec:experimental}

Let us provide a rough estimate of the characteristic parameters. 
Assuming a uniform rubidium BEC with a density of $\rho_{\rm c} = 10^{14}\,\rm{cm}^{-3}$ and the  
decay constant $D\sim 5.8\times 10^{-30}\,\rm{cm}^6\rm{s}^{-1}$ \cite{Burt:1997coh} 
yields $\gamma = 3D\rho_{\rm c}^2 \sim 0.2 \,\rm{s}^{-1}$.
Similarly, a ytterbium BEC with $D \sim 4\times 10^{-30}\,\rm{cm}^6\rm{s}^{-1}$ \cite{Takuso:2003spin} 
and the same density gives $\gamma \sim 0.1 \,\rm{s}^{-1}$. 
Making the condensates more dilute $\rho_{\rm c} = 10^{13}\,\rm{cm}^{-3}$ reduces $\gamma$ by two orders of 
magnitude and leads to longer inverse damping times between $500\,\rm{s}$ and $1000\,\rm{s}$.
These time scales limit the life-time of the condensate itself and of the phononic excitations, 
but do also pose restrictions on quantum enhanced sensing. 
Note that the noise rate $\gamma^+$ is even larger than the above values of $\gamma$.

\subsection{Comparison to Landau \& Beliaev damping}\label{sec:landau}

Let us compare the above values to other decoherence and damping channels often discussed in BECs,
i.e., Landau and Beliaev damping.
Landau damping corresponds to a process where two 
quasi-particle excitations (i.e., atoms in excited motional states) 
interact such that their energy and momentum are combined into a single higher energetic 
quasi-particle leaving the remaining atom as a condensate atom. 
It was initially discussed in \cite{Szepfalusy:1974on,Shi:1998fin,Fedichev:1998damp}. 
An expression for the damping constant $\gamma$ in a uniform BEC for general temperatures was derived in 
\cite{Pitaevskii:1997land,Giorgini:1998dam}. 
As one would expect, Landau damping is suppressed for low temperatures $T$ (where only a few 
excitations are present) and scales with $T^4$. 
Furthermore, it grows linearly with the quasi-particle momentum. 
Assuming a low temperature of 200~pK and a dilute and fairly large (elongated) BEC with size 200~$\mu$m
and density $\rho_{\rm c} = 10^{13}\,\rm{cm}^{-3}$, the associated damping rate $\gamma$ can be suppressed 
down to $10^{-5}~\rm{s}^{-1}$. 
In this regime, it is negligible in comparison with the three-body-loss -- unless very high 
quasi-particle momenta are considered. 

Beliaev damping corresponds to the scattering of 
a quasi-particle excitation and a condensate atom leading to two 
quasi-particle excitations that share the kinetic energy and the momentum of the initial excitation. 
Because it does not require a second quasi-particle excitation, it can also exist at zero temperature -- 
but it scales with the fifth power of the quasi-particle momentum \cite{Giorgini:1998dam}. 
Thus, for low momenta, Beliaev damping can be suppressed down to rates $10^{-9}~\rm{s}^{-1}$ 
(i.e., even more strongly than Landau damping), but it can become dominant for larger momenta. 

For higher temperatures and/or larger quasi-particle momenta, Landau and Beliaev damping start 
to dominate in comparison to three-body loss. 
However, their consequences are completely analogous to the results discussed above:
On the linearized quasi-particle level, their impact is described by the same effective master 
equation~\eqref{eq:masterunif}, just with adapted decay rates $\gamma_\pm$, see also \cite{Howl_2017} and Appendix \ref{app:bellanddamp}. 

It should be stressed here that the above considerations are based on the linearized quasi-particle picture.
If too many quasi-particles are present, higher-order non-linearities may become important and induce 
additional effects such as de-phasing.

\subsection{Example Application: Gravity Sensing}\label{sec:example}

To give an intuition for the effect of decoherence in an experimental situation, we will present
the sensing of oscillating gravitational fields as an example application in the following. 
Note that it is not our intention to give a full experimental proposal, 
which requires several steps:
First, one has to make sure that the oscillating gravitational field changes the quantum state of the probe 
(i.e., the condensate) enough to be detectable in principle. 
Second, one has to find a way to actually detect this change experimentally. 
Third, one has to distinguish the measured signal from noise and background processes. 
Here, we mainly focus on the first step -- if it cannot be accomplished successfully, 
there is no need to consider the further steps.

In \cite{ratzel_dynamical_2018}, it has been theoretically investigated how the time-dependent
gravitational field of a small oscillating gold sphere acts on a nearby BEC. 
In particular, it has been shown that
phonon modes respond to the gravitational field, 
especially if the oscillation of the gravitational acceleration is on resonance with a phonon mode. 
In that case (direct driving), the interaction between the phonon mode and the gravitational field 
can be represented by a displacement Hamiltonian $\hat H_D$.  
Starting in the ground state (i.e., no initial squeezing), the displacement is detectable 
(in principle) if it corresponds to an expectation value of the phonon number operator 
(in this mode) of order unity\footnote{Still, measuring a single phonon in a BEC is a non-trivial 
task, see, e.g., the Conclusions section of \cite{ratzel_dynamical_2018} or 
[Detection Scheme for Acoustic Quantum Radiation in Bose-Einstein Condensates,
Ralf Sch\"utzhold,
Phys.\ Rev.\ Lett.\ {\bf 97}, 190405 (2006)].}.

As an explicit example, let us consider a cylindrical ytterbium BEC of length $L$ and diameter $d$, 
which is axially oriented with respect to the source mass' motion. 
To simplify the comparison, we consider the same parameters as discussed in \cite{ratzel_dynamical_2018}.
Thus, we assume a rather large BEC, where $L$ and $d$ both are approximately $300\,\mathrm{\mu m}$, 
containing $2\times 10^8$ condensed atoms.
This corresponds to a rather low density of $10^{13}\,\mathrm{cm}^{-3}$, 
for which the damping rates have already been discussed above. 
If we consider a gold sphere of $200\,$g, oscillating with a frequency of $8\,$Hz and an amplitude of 
$2\,$mm at an average distance of $3\,$mm from the source mass's surface to the BEC, the impact of 
the gravitational field on the phonon mode with the wave vector $\mathbf{k}=(10\pi/L,0,0)$
pointing in the axial direction, would lead to a detectable displacement after $10\,$s. 

For the displacement scheme, the fundamental uncertainty for a measurement is given by 
equation~(\ref{eq:displacement-decay}).  
More explicitly, the change of the displacement is proportional to the source mass. 
Thus, in the absence of decoherence, an initial squeezed state implies a scaling of the minimum 
detectable mass $\Delta m$ as $e^{-r}$. 
For example, for $r=1$, $r=2$ and $r=5$ and the system parameters above, 
this corresponds to a detectable mass of about $70\,$g, $30\,$g and $1\,$g, 
respectively.  

In the ideal case, a squeezing of $r=5$ corresponds to an increase in sensitivity of more than 
two orders of magnitude. 
Note that this corresponds to a mean phonon number of 
$\langle\hat n\rangle=\sinh^2(r)\approx 5.5\times10^3$, i.e., a highly excited state 
(with a correspondingly large energy).
Apart from the experimental challenges to actually create (and detect) such a state, 
one should carefully scrutinize the underlying approximations for such a highly excited state,
e.g., regarding the impact of non-linearities (which might induce de-phasing etc.). 

Moreover, even with the small value of $\gamma\sim10^{-3}~\rm{s}^{-1}$, 
leading to $\gamma_\mathbf{k}^+\sim 10^{-2}~\rm{s}^{-1}$ for the specific mode under consideration,
the quantum enhancement of the sensitivity would already be decreased to a factor $4$ after $10\,$s.
This corresponds to a detectable mass of $50\,$g (instead of $1\,$g in the ideal case).

As the change of displacement is proportional to the driving time $t$ (see \cite{ratzel_dynamical_2018}) 
while the decrease in sensitivity due to decoherence is proportional to the square root of $t$
(for large squeezing), the sensitivity still increases with driving time -- provided that the 
resonance condition can be maintained during that time and that the BEC is not disturbed too much.

\bigskip

\section{Conclusions and Discussions}\label{sec:conclusions}

For the phonon modes of Bose-Einstein condensates in ultra-cold atomic vapor, we studied the impact of 
decoherence caused by the omnipresent process of three-body loss on quantum enhanced sensing. 
In contrast to other effects such as Landau and Beliaev damping, this process cannot be suppressed by 
going to ultra-low temperatures or energies. 
Three-body loss can be suppressed by making the condensate more dilute 
or by reducing the atomic interaction strength \cite{Moerdijk:1996dec,fedichev1996three}, 
but this would also diminish other important quantities such as the speed of sound or the total number of 
atoms in the BEC. 
As a result, the overall dynamics of the BEC would become slower or the available phase 
space (e.g., possible amount of squeezing and maximum number of phonons) would shrink -- 
which poses additional challenges for sensing schemes.

Another way of suppressing three-body loss would be to confine the BEC spatially, i.e., 
to make it effectively lower dimensional (see Appendix \ref{sec:lowerdim} for a detailed discussion). 
In that case, a reduction of dimension to 1d can lead to a suppression
of the loss rate by several orders of magnitude for ultra-low temperatures in the weakly interacting 
Bogoliubov regime \cite{tolra_observation_2004,haller_three-body_2011,kormos_exact_2011}
when the atomic density is kept constant, implying the corresponding reduction
of the number of atoms. 
In the strongly interacting Tonks-Girardeau regime, the suppression
is much more significant and three-body loss can be reduced very strongly.
However, the restriction to the Tonks-Girardeau regime leads to an upper bound
on the atomic density for a fixed 1d scattering length \cite{kormos_exact_2011}. 
This implies an upper bound on the number of atoms
that can be confined in the same trap.

It has been shown that controlling the environment may lead to a partial retrieval of quantum enhancement 
in sensing \cite{Albarelli2018restoringheisenberg,Bai:2019retr}. 
Therefore, the effect of three-body loss might be limited by precisely monitoring dimer molecules and 
excess atoms (see also \cite{schutzhold2010quantum}) even though this is experimentally very challenging. 

Depending on the squeezing parameter $r$ employed for quantum enhanced sensing, we found a rapid 
deterioration of precision on a time scale $t=\ord(e^{-2r}/ \gamma^+)$, which is much faster than the 
decay of phonons on the time scale $t=\ord(1/ \gamma^-)$.  
This hierarchy of time scales is analogous to the difference between relaxation time $T_1$ and 
coherence time $T_2$ known from quantum information theory, for example. 

In principle, one could counteract the decoherence induced growth of uncertainty by permanent squeezing. 
However, the required squeezing rate $\Xi$ would be quite large $\Xi=\ord(e^{2r} \gamma^+)$ and thus 
experimentally challenging.
In addition, since this squeezing operation should demagnify the direction $\mathbf{u}_\lambda$ 
corresponding to the small eigenvalue $\lambda_-=e^{-2r}$ of the covariance matrix, i.e., 
precisely the same direction as the signal to be measured, 
there is the danger to reduce the signal as well, which would be counterproductive. 
This can be understood by inspecting Fig.~\ref{figure}, for example.
In order to enhance the sensitivity for measuring a displacement in $X$ direction,
one would squeeze the initial state in that direction as well. 
However, in order to counteract the decoherence induced growth of uncertainty, one would also 
have to squeeze the final state in the same direction, which would also demagnify the 
signal to be measured. 
In addition, such a squeezing would also generate further excitations, 
i.e., a larger number of phonons, which pose further problems.

We would also like to stress that we considered general bounds on the achievable accuracy based on the 
behavior of the variances or the Quantum Fisher Information~\eqref{eq:QFI}, but we did not provide a 
concrete measurement scheme (which can also be quite challenging experimentally). 
Nevertheless, our general limits on the measurement time $t$ and the squeezing $r$ should be taken into 
account for proposals involving quantum enhanced sensing applications based on BECs.

\acknowledgments

We thank Richard Howl for useful remarks and discussions.
DR acknowledges funding by the Marie Skłodowska-Curie Action IF program -- Project-Name "Phononic Quantum Sensors for Gravity" (PhoQuS-G) -- Grant-Number 832250.
RS acknowledges funding by the Deutsche Forschungsgemeinschaft (DFG, German Research Foundation) --  
Project-ID 278162697 -- SFB 1242.


\bibliographystyle{unsrt}
\bibliography{r2-bibliography}

\onecolumngrid
\appendix

\section{Time evolution of displacement vector and covariance matrix}

In equation (\ref{eq:masterunif}), we found that, in the rotating-wave approximation, the time evolution of the density matrix of the field $\hat\chi$ is given by the master equation
\bea
\frac{d\hat\varrho_{\chi}}{dt}
&=& 
-i\left[\hat H_0 ,\hat\varrho_{\chi}\right] 
+  \sum_I \gamma_{I}^u \Big( \hat{b}_I \hat\varrho_{\chi} \hat{b}_I^\dagger - 
\frac12 \left\{  \hat{b}_I^\dagger  \hat{b}_I  ,\hat\varrho_{\chi} \right\} \Big)  
+  \sum_I \gamma_{I}^v \Big( \hat{b}_I^\dagger \hat\varrho_{\chi} \hat{b}_I 
- \frac12 \left\{  \hat{b}_I  \hat{b}_I^\dagger  ,\hat\varrho_{\chi} \right\} \Big)   \,.
\ea
With $\hat{H}_0 = \sum_I \omega_I \hat{b}^\dagger_I \hat{b}_I$ and $\gamma_I^\pm = \gamma_{I}^u \pm \gamma_{I}^v$, the time evolution of expectation values of first order and second order operators becomes
\bea
\nonumber \frac{d\langle \hat{b}_I\rangle }{dt} &=& -i\rm{Tr}\left(\hat{b}_I\left[\hat H_0,\hat\varrho_\chi\right]\right)  + \gamma_{I}^u\,\rm{Tr}\Bigg(\hat{b}_I\Bigg( \hat b_I \hat\varrho_\chi\hat b_I^\dagger -\frac12 \left\{ \hat b_I^\dagger \hat b_I  ,\hat\varrho_\chi \right\} \Bigg)\Bigg) + \gamma_{I}^v\,\rm{Tr}\Bigg(\hat{b}_I\Bigg( \hat b_I^\dagger \hat\varrho_\chi \hat b_I  -\frac12 \left\{ \hat b_I \hat b_I^\dagger  ,\hat\varrho_\chi \right\} \Bigg)\Bigg)\\
\nonumber &=& -i\omega_I \langle\hat{b}_I\rangle + \gamma_{I}^u\, \langle \hat b_I^\dagger \hat{b}_I^2 - (\hat{b}_I \hat b_I^{\dagger} \hat{b}_I + \hat b_I^\dagger \hat{b}_I^2 )/2\rangle   + \gamma_{I}^v\langle  \hat{b}_I^2 \hat b_I^\dagger  - (\hat{b}_I^2 \hat b_I^{\dagger}  + \hat{b}_I \hat b_I^\dagger \hat{b}_I )/2\rangle \\
&=&  - (i\omega_I + \gamma_I^-/2) \langle\hat{b}_I\rangle  \,,
\eea
\bea
\nonumber \frac{d\langle \hat{b}_I^2\rangle }{dt} &=& -i\rm{Tr}\left(\hat{b}_I^2\left[\hat H_0,\hat\varrho_\chi\right]\right) + \gamma_{I}^u \,\rm{Tr}\Bigg(\hat{b}_I^2\Bigg( \hat b_I  \hat\varrho_\chi \hat b_I^\dagger  - \frac12 \left\{ \hat b_I^\dagger \hat b_I  ,\hat\varrho_\chi \right\} \Bigg)\Bigg) + \gamma_{I}^v \,\rm{Tr}\Bigg(\hat{b}_I^2\Bigg( \hat b_I^\dagger \hat\varrho_\chi  \hat b_I  -\frac12 \left\{ \hat b_I \hat b_I^\dagger  ,\hat\varrho_\chi \right\} \Bigg)\Bigg)\\
\nonumber &=& -2i\omega_I \langle\hat{b}_I\rangle + \gamma_{I}^u\, \langle \hat b_I^\dagger\hat{b}_I^3  - (\hat{b}_I^2 \hat b_I^{\dagger} \hat{b}_I + \hat b_I^\dagger \hat{b}_I^3 )/2\rangle  + \gamma_{I}^v\,\langle  \hat{b}_I^3 \hat b_I^\dagger  - (\hat{b}_I^3 \hat b_I^{\dagger}  + \hat{b}_I \hat b_I^\dagger \hat{b}_I^2 )/2\rangle \\
&=&  - (2i\omega_I + \gamma_I^-) \langle\hat{b}_I^2\rangle  \,,
\eea
\bea
\nonumber \frac{d\langle \hat{b}_I^\dagger \hat{b}_I\rangle }{dt} &=& -i\rm{Tr}\left(\hat{b}_I^\dagger \hat{b}_I\left[\hat H_0,\hat\varrho_\chi\right]\right)\\
\nonumber &&  + \gamma_{I}^u\,\rm{Tr}\Bigg(\hat{b}_I^\dagger \hat{b}_I \Bigg( \hat b_I  \hat\varrho_\chi \hat b_I^\dagger  - \frac12 \left\{ \hat b_I^\dagger   \hat b_I  ,\hat\varrho_\chi \right\} \Bigg)\Bigg)  + \gamma_{I}^v\,\rm{Tr}\Bigg(\hat{b}_I^\dagger \hat{b}_I \Bigg( \hat b_I^\dagger \hat\varrho_\chi \hat b_I - \frac12 \left\{  \hat b_I  \hat b_I^\dagger ,\hat\varrho_\chi \right\} \Bigg)\Bigg)\\
\nonumber &=& \gamma_{I}^u\,\langle \hat b_I^{\dagger 2} \hat{b}_I^2  - \hat{b}_I^\dagger \hat{b}_I \hat b_I^{\dagger} \hat{b}_I \rangle + \gamma_{I}^v\langle \hat{b}_I \hat{b}_I^\dagger \hat{b}_I \hat b_I^\dagger  - (\hat{b}_I^\dagger \hat{b}_I^2 \hat b_I^{\dagger}  + \hat{b}_I \hat b_I^{\dagger 2} \hat{b}_I )/2\rangle\\
&=& -\gamma_I^- \langle \hat{b}_I^\dagger \hat{b}_I \rangle  + \gamma_{I}^v  \\
 \frac{d\langle  \hat{b}_I \hat{b}_I^\dagger \rangle }{dt} &=& \frac{d\langle \hat{b}_I^\dagger \hat{b}_I\rangle }{dt} =  -\gamma_I^- \langle \hat{b}_I^\dagger \hat{b}_I \rangle  + \gamma_{I}^v = -\gamma_I^- \langle \hat{b}_I \hat{b}_I^\dagger \rangle  + \gamma_{I}^u \,.
\ea
and for $n\neq m$
\bea
\nonumber \frac{d\langle \hat{b}_I^\dagger \hat{b}_J\rangle }{dt} &=&  - (i(\omega_J - \omega_I) + (\gamma_I^- + \gamma_J^- )/2) \langle\hat{b}_I^\dagger \hat{b}_J \rangle  \\
\nonumber \frac{d\langle \hat{b}_I \hat{b}_J\rangle }{dt} &=&  - (i(\omega_J + \omega_I) + (\gamma_I^- + \gamma_J^- )/2) \langle\hat{b}_I \hat{b}_J \rangle\,.
\eea
Defining the rotating frame 
\bea
	\overline{\langle \hat{b}_I\rangle} &=& \langle \hat{b}_I\rangle e^{i\omega_I t} \\
	\overline{\langle \hat{b}_I^2\rangle} &=& \langle \hat{b}_I\rangle e^{2i\omega_I  t} \\
	\overline{\langle \hat{b}_I^\dagger\hat{b}_I\rangle} &=& \langle \hat{b}_I^\dagger\hat{b}_I\rangle\\
	\overline{\langle \hat{b}_I^\dagger\hat{b}_J\rangle} &=& \langle \hat{b}_I^\dagger\hat{b}_J\rangle e^{i (\omega_J  - \omega_I) t}\\
	\overline{\langle \hat{b}_I\hat{b}_J\rangle} &=& \langle \hat{b}_I\hat{b}_J\rangle e^{i(\omega_J + \omega_I ) t}\,,
\eea
the differential equations become
\bea
\nonumber \frac{d \overline{\langle \hat{b}_I\rangle} }{dt} &=& -\frac{\gamma_I^-}{2} \overline{\langle\hat{b}_I\rangle}   \\
\frac{d \overline{\langle \hat{b}_I^2\rangle} }{dt} &=& - \gamma_I^- \overline{\langle\hat{b}_I^2\rangle}   \\
\nonumber \frac{d \overline{\langle \{ \hat{b}_{I}^{\dagger}, \hat{b}_{I} \}_+ \rangle} }{dt} &=&  -\gamma_I^- \overline{\langle \{ \hat{b}_{I}^{\dagger}, \hat{b}_{I} \}_+ \rangle} + \gamma_I^+  \,,
\eea
and for $n\neq m$
\bea
	\frac{d \overline{\langle \hat{b}_I^\dagger \hat{b}_J\rangle} }{dt} &=&  - \frac{\gamma_I^- + \gamma_J^- }{2} \overline{\langle\hat{b}_I^\dagger \hat{b}_J \rangle}  \\
	\frac{d \overline{\langle \hat{b}_I \hat{b}_J\rangle} }{dt} &=&  - \frac{\gamma_I^- + \gamma_J^- }{2} \overline{\langle\hat{b}_I \hat{b}_J \rangle}\,.
\eea
The solutions to these differential equations are easily found as
\bea
	\overline{\langle \hat{b}_I\rangle} &=& e^{-\gamma_I^- t/2} \overline{\langle \hat{b}_I\rangle_0} \\
	\overline{\langle \hat{b}_I^2\rangle} &=& e^{- \gamma_I^- t} \overline{\langle \hat{b}_I^2\rangle_0}  \\
	\overline{\langle \{ \hat{b}_I^{\dagger}, \hat{b}_I \}_+ \rangle} &=& e^{-\gamma_I^- t} (\overline{\langle \{ \hat{b}_I^{\dagger}, \hat{b}_I \}_+ \rangle_0} - \gamma_I^+ / \gamma_I^-) + \gamma_I^+ / \gamma_I^-  \,,
\eea
and for $n\neq m$
\bea
	\overline{\langle \hat{b}_I^\dagger \hat{b}_J \rangle} &=& e^{-(\gamma_I^- + \gamma_J^- )t/2} \overline{\langle \hat{b}_I^\dagger \hat{b}_J \rangle_0} \\
	\overline{\langle \hat{b}_I \hat{b}_J \rangle} &=& e^{-(\gamma_I^- + \gamma_J^- )t/2} \overline{\langle \hat{b}_I\hat{b}_J \rangle_0} \,.
\eea
We define the operator valued vector $\hat{\xi} = (\hat{b}_1,\hat{b}_1^\dagger, ... , \hat{b}_I, \hat{b}_I^\dagger, ... )^\rm{Tp}$ containing all creation and 
annihilation operators of the quasi-particle field $\hat{\chi}$ and $\hat{\xi}^\dagger = (\hat{b}_1^\dagger,\hat{b}_1, ... , \hat{b}_I^\dagger, \hat{b}_I, ... )^\rm{Tp}$.
In this basis, the time evolution of the displacement vector and covariance matrix become in the co-rotating frame
\bea
	\label{eq:Dt} \mbf{D}(t) &=& \mbf{R}^*(t)\cdot \langle \hat{\xi} \rangle_t = e^{- \mbf{\Gamma}_- t/2} \cdot \mbf{D}(0)\, \\
	 \label{eq:Sigmat}  \mbf{\Sigma}(t) &=& \mbf{R}^*(t) \cdot  \left\langle \left\{ \hat{\xi} - \mbf{D}(t) , (\hat{\xi} - \mbf{D}(t))^{\dagger \rm{Tp} }\right\}_+ \right\rangle_t \cdot  \mbf{R}(t) \\
	\nonumber &=&  e^{- \mbf{\Gamma}_- t/2} \cdot  \left(  \mbf{\Sigma}(0) - \mbf{\Sigma}_\infty \right)\cdot  e^{- \mbf{\Gamma}_- t/2}  + \Sigma_\infty\,,
\eea
where $\mbf{\Sigma}_\infty =\mbf{\Gamma}_+ \cdot \mbf{\Gamma}_-^{-1}$ and the diagonal matrices $\mbf{\Gamma}_\pm$ and $\mbf{R}(t)$ are defined as $\left(\mbf{\Gamma}_\pm\right)_{2I-1} = \left(\mbf{\Gamma}_\pm\right)_{2I} = \gamma_I^\pm$ and $\mbf{R}(t)_{2I-1}=e^{-i\omega_I t}$, $\mbf{R}(t)_{2I}=e^{i\omega_I t}$, respectively. Equation (\ref{eq:Sigmat}) is the solution of equation (\ref{eq:oem-rotating-wave}) in the main text.

\section{Parameter estimation with a squeezed vacuum state}

For a single mode squeezed state
\bea
	\hat{\varrho}_\chi(0) = \hat{S}_I(\zeta)\hat{\varrho}_{\chi,\rm{vac}} \hat{S}_I(\zeta)^\dagger  \,, 
\eea 
where $\hat{S}_I(\zeta)=\exp((\zeta^*\hat{b}_I^2 - \zeta \hat{b}_I^{\dagger 2})/2)$ is the squeezing operator with $\zeta = r e^{2i\theta}$, we have 
\bea
	\mbf{\Sigma}(t)_I &=& e^{-\gamma_I^- t}\mbf{S}_I(\zeta) \cdot \mbf{S}_I(\zeta)  + \frac{\gamma_I^+}{\gamma_I^-}(1 - e^{-\gamma_I^- t} )\mathbb{I}
\eea
where 
\bea
	\mbf{S}_I(\zeta) = \left( \begin{array}{cc}  \cosh(r)   & - e^{2i\theta} \sinh(r) \\
												- e^{-2i\theta} \sinh(r)   &  \cosh(r) 
							 \end{array} \right)\,.
\eea
is the single mode squeezing operator's action on mode $I$.  Without loss of generality, we assume $\theta = 0$ in the following.

\subsection{Estimation of displacement}
\label{sec:estdispl}

We consider an additional displacement $\hat{D}_I(\mu)$, where $\mu = |\mu| e^{i\varphi_\mu}$. This does not change the covariance matrix and, for a squeezed vacuum state, the displacement vector simply becomes
\bea
	\mbf{D}(t)_{I,\mu}  & = &   (\mu, \mu^*)^\rm{Tp}   \,.
\eea
The quantum Fisher information for the estimation of $|\mu|$ is then given as
\bea
	F_{|\mu|}(t) &=& 2\left(\mbf{D}(t)_{I,\mu}^\prime \right)^{*\rm{Tp}} \cdot \mbf{\Sigma}_I(t)^{-1} \cdot \mbf{D}(t)_{I,\mu}^\prime \,,
\eea
where
\bea
	\mbf{D}(t)_{I,\mu}^\prime = \frac{d}{d\epsilon}\mbf{D}(t)_{I,(|\mu|+\epsilon)e^{i\varphi_\mu}} \Big|_{\epsilon=0} =   ( e^{i\varphi_\mu} , e^{-i\varphi_\mu})^\rm{Tp} \,.
\eea
Again, without loss of generality, we assume $\theta = 0$. In the eigenbasis of $\mbf{\Sigma}(t)_I$, we find
\bea
	\mbf{D}(t)_{I,\mu}^{\prime\,\rm{eb}} &=&  \sqrt{2} (\cos(\varphi_\mu) , -i \sin(\varphi_\mu))^\rm{Tp} 	\,
\eea
and we find that the QFI is maximized for $\varphi_\mu = 0$ leading to
\bea
	F_{|\mu|}(t) &=& 4\lambda_-(t)^{-1} \,.
\eea
We find that quantum enhanced sensing of the amplitude of displacement 
with a squeezed vacuum state has a sensitivity that scales at most with $1/\sqrt{\langle \hat{n}\rangle}$.

Let us assume that we want to estimate the phase of a displacement. In that case,
\bea
	\mbf{D}(t)_{I,\mu}^\prime = \frac{d}{d\epsilon}\mbf{D}(t)_{I,|\mu|e^{i(\varphi_\mu+\epsilon)}} \Big|_{\epsilon=0} =   |\mu|( i e^{i\varphi_\mu} , -i e^{-i\varphi_\mu})^\rm{Tp} 
\eea
and 
\bea
	\mbf{D}(t)_{I,\mu}^{\prime\,\rm{eb}} &=&  -\sqrt{2} |\mu| (\sin(\varphi_\mu) , i \cos(\varphi_\mu))^\rm{Tp} .	\,
\eea
We find that the QFI is maximized for $\varphi_\mu = \pi/2$ leading to
\bea
	F_{|\mu|}(t) &=& 4|\mu|^2 \lambda_-(t)^{-1} \,.
\eea
Now, the corresponding sensitivity 
scales with the product of the square root of the number of phonons in the initial state
and the square root of the number of phonons $|\mu|^2$ created by the signal. 
This can be associated with Heisenberg scaling if the number of squeezed 
phonons $\sinh^2(r)$ is of the same order as $|\mu|^2$ and
a small signal is enhanced to achieve the amplitude $|\mu|$.

\subsection{Estimation of a rotation angle}
\label{sec:estrot}

We apply an additional rotation $\hat{R}_I(\vartheta_\epsilon) = \exp(-i\vartheta_\epsilon \hat{b}_I^\dagger \hat{b}_I)$ of a Gaussian state acts on the covariance matrix as
\bea
	\nonumber \mbf{\Sigma}(t)_{I,\vartheta_\epsilon} & = &  \mbf{R}_I(\vartheta_\epsilon) \cdot \mbf{\Sigma}(t)_I \cdot \mbf{R}_I(\vartheta_\epsilon)^{*\rm{Tp}}   \,.
\eea
where $\mbf{R}_I(\vartheta_\epsilon)=\rm{diag}(e^{-i\vartheta_\epsilon},e^{i\vartheta_\epsilon})$.
For the QFI, we have
\bea
	F_{\vartheta_\epsilon}(t) &=& \frac{1}{2(1 + P(t)^2)}\rm{Tr}\left[\left(\mbf{\Sigma}(t)_{I,\vartheta_\epsilon}^{-1} \mbf{\Sigma}(t)_{I,\vartheta_\epsilon}^\prime\right)^2 \right]   \,,
\eea
where
\bea
	 \mbf{\Sigma}(t)_{I,\vartheta_\epsilon}^{-1} &=&  \mbf{R}_I(\vartheta_\epsilon) \cdot \mbf{\Sigma}(t)_I^{-1} \cdot \mbf{R}_I(\vartheta_\epsilon)^{*\rm{Tp}} \\
	 \nonumber \mbf{\Sigma}(t)_{I,\vartheta_\epsilon}^\prime &=& \frac{d}{d\vartheta_\epsilon}\mbf{\Sigma}(t)_{I,\vartheta_\epsilon} =  \mbf{\Omega} \cdot \mbf{R}_I(\vartheta_\epsilon) \cdot \mbf{\Sigma}(t)_I \cdot \mbf{R}_I(\vartheta_\epsilon)^{*\rm{Tp}}  -   \mbf{R}_I(\vartheta_\epsilon) \cdot \mbf{\Sigma}(t)_I \cdot \mbf{R}_I(\vartheta_\epsilon)^{*\rm{Tp}} \cdot \mbf{\Omega} \\
	 &=& \mbf{\Omega} \cdot \mbf{\Sigma}(t)_{I,\vartheta_\epsilon} - \mbf{\Sigma}(t)_{I,\vartheta_\epsilon} \cdot \mbf{\Omega} 
\eea
and $\Omega = \rm{diag}(-i,i)$. Also, we find
\bea
	 \rm{Tr}\left[\left(\mbf{\Sigma}(t)_{I,\vartheta_\epsilon}^{-1} \mbf{\Sigma}(t)_{I,\vartheta_\epsilon}^\prime\right)^2 \right] &=& \rm{Tr}\left[ \left( \mbf{\Sigma}(t)_I^{-1} \left(\mbf{\Omega} \cdot \mbf{\Sigma}(t)_I - \mbf{\Sigma}(t)_I \cdot \mbf{\Omega}\right) \right)^2 \right]\,,
\eea
which means that the estimation of phase with a squeezed state does not depend on the base point $\vartheta_\epsilon$. In the eigenbasis of 
$\mbf{\Sigma}(t)_I$, we find
\bea
	\mbf{\Omega}^\rm{eb} &=&  i \left(\begin{array}{cc} 0 & 1 \\ 1  & 0 \end{array}\right) \quad\rm{and}\\
	\left(\mbf{\Omega} \cdot \mbf{\Sigma}(t)_I - \mbf{\Sigma}(t)_I \cdot \mbf{\Omega}\right)^\rm{eb} &=& i(\lambda_-(t) - \lambda_+(t)) \left(\begin{array}{cc} 0 & -1 \\ 1  & 0 \end{array}\right)  \\
	\mbf{\Sigma}(t)_I^{-1\,\rm{eb}} \left(\mbf{\Omega} \cdot \mbf{\Sigma}(t)_I - \mbf{\Sigma}(t)_I \cdot \mbf{\Omega}\right)^\rm{eb} &=& i(\lambda_-(t) - \lambda_+(t)) \left(\begin{array}{cc} 0 & -\lambda_-(t)^{-1} \\ \lambda_+(t)^{-1}  & 0 \end{array}\right)	\,. 
\eea
Therefore,
\bea
	\rm{Tr}\left[ \left( \mbf{\Sigma}(t)_I^{-1} \left(\mbf{\Omega} \cdot \mbf{\Sigma}(t)_I - \mbf{\Sigma}(t)_I \cdot \mbf{\Omega}\right) \right)^2 \right]  = 2\frac{(\lambda_-(t) - \lambda_+(t))^2}{\lambda_-(t)\lambda_+(t)}\,.
\eea
We obtain for the QFI\footnote{{We recover the result of \cite{pinel2013quantum} equation (16) with $\alpha=0$, $\lambda_-(t) = 1/(P_0 \sigma^2)$ and $\lambda_+(t) = \sigma^2/P_0$.}}
\bea\label{eq:QFIrotation}
	F_{\vartheta_\epsilon}(t) &=& \frac{(\lambda_-(t) - \lambda_+(t))^2}{\lambda_-(t)\lambda_+(t) + 1} \,.
\eea
For $e^{-2r}\ll 1$ and $\gamma_I^\pm t\ll 1$, we obtain
\bea
	\Delta \vartheta_\epsilon = \frac{1}{\sqrt{F_{\vartheta_\epsilon}}} \approx  e^{-2r} \sqrt{ 2 + e^{2r}\gamma^+t} \,.
\eea

\subsection{Estimation of the squeezing amplitude}
\label{sec:estsqueezing}

Now, let us apply an additional squeezing $\hat{S}_I(\nu)$, where $\nu = s e^{2i\phi_\nu}$ and bound the sensitivity for estimating the parameter $s$.
The action on the covariance matrix is
\bea
	\mbf{\Sigma}(t)_{I,\nu} & = &  \mbf{S}_I(\nu) \cdot \mbf{\Sigma}(t)_I \cdot \mbf{S}_I(\nu)    \,.
\eea
The quantum Fisher information for the estimation of $s=|\nu|$ is 
\bea
	F_{s}(t) &=& \frac{1}{2(1 + P_\nu(t)^2)}\rm{Tr}\left[\left(\mbf{\Sigma}(t)_I^{-1} \cdot \mbf{\Sigma}(t)_{I,\nu}^\prime\right)^2 \right]   \,,
\eea
where  
\bea
	 \nonumber \mbf{\Sigma}(t)_{I,\nu}^\prime = \frac{d}{d\epsilon}\mbf{\Sigma}_{I,\epsilon e^{i\varphi_\nu}}(t) \Big|_{\epsilon = 0} &=&  \bar{\mbf{Y}}(\phi_\nu) \cdot \mbf{\Sigma}(t)_I \cdot \mbf{S}_I(\nu) + \mbf{S}_I(\nu) \cdot \mbf{\Sigma}(t)_I\cdot \bar{\mbf{Y}}(\phi_\nu) \\
	&=&  \bar{\mbf{Y}}(\phi_\nu) \cdot \mbf{S}_I(\nu)^{-1} \cdot \mbf{\Sigma}(t)_{I,\nu}  +  \mbf{\Sigma}(t)_{I,\nu}\cdot  \mbf{S}_I(\nu)^{-1} \cdot \bar{\mbf{Y}}(\phi_\nu)
\eea
and
\bea
	\bar{\mbf{Y}}(\phi_\nu) = \left.\frac{d}{ds} \mbf{S}(s e^{i\phi_\nu})\right|_{s=0} = \left(\begin{array}{cc} \sinh(s) & -e^{2i\phi_\nu}\cosh(s) \\ -e^{-2i\phi_\nu}\cosh(s)  & \sinh(s) \end{array}\right)\,.
\eea
We can rewrite the QFI as
\bea
	\nonumber F_{\phi_\nu}(t) &=& \frac{1}{(1 + P_\nu(t)^2)} \left( \rm{Tr}\left[\left( \bar{\mbf{Y}}(\phi_\nu) \cdot \mbf{S}_I(\nu)^{-1} \right)^2 \right]  + \rm{Tr}\left[  \bar{\mbf{Y}}(\phi_\nu) \cdot \mbf{S}_I(\nu)^{-1} \cdot  \mbf{\Sigma}(t)_{I,\nu} \cdot \mbf{S}_I(\nu)^{-1} \cdot \bar{\mbf{Y}}(\phi_\nu) \cdot \mbf{\Sigma}(t)_{I,\nu}^{-1} \right] \right) \\
	&=& \frac{1}{(1 + P_\nu(t)^2)} \left( 2 + \rm{Tr}\left[  \mbf{S}_I(\nu)^{-1} \cdot \bar{\mbf{Y}}(\phi_\nu) \cdot  \mbf{\Sigma}(t)_I \cdot  \bar{\mbf{Y}}(\phi_\nu) \cdot \mbf{S}_I(\nu)^{-1} \cdot \mbf{\Sigma}(t)_I^{-1} \right] \right)  \,.
\eea
We transform into the 
eigenbasis of $\mbf{\Sigma}(t)_I$ such that
\bea
	\mbf{\Sigma}(t)_I^\rm{eb} &=& e^{-\gamma_I^- t}\left(\begin{array}{cc} e^{-2r} & 0 \\ 0  & e^{2r} \end{array}\right) + \frac{\gamma_I^+}{\gamma_I^-}(1 - e^{-\gamma_I^- t} )\mathbb{I} =: \left(\begin{array}{cc} \lambda_-(t) & 0 \\ 0  & \lambda_+(t) \end{array}\right)\,, \\
	\bar{\mbf{Y}}(\phi_\nu)^\rm{eb} \cdot \mbf{S}_I(\nu)^{-1\,\rm{eb}} &=& \mbf{S}_I(\nu)^{-1\,\rm{eb}} \cdot  \bar{\mbf{Y}}(\phi_\nu)^\rm{eb}  = \left(\begin{array}{cc} -\cos(2\phi_\nu) & -i\sin(2\phi_\nu) \\ i\sin(2\phi_\nu)  & \cos(2\phi_\nu) \end{array}\right) \,, 
\eea
and
\bea
	\nonumber && \mbf{S}_I(\nu)^{-1} \cdot \bar{\mbf{Y}}(\phi_\nu) \cdot  \mbf{\Sigma}(t)_I \cdot  \bar{\mbf{Y}}(\phi_\nu) \cdot \mbf{S}_I(\nu)^{-1} \cdot \mbf{\Sigma}(t)_I^{-1} \\
	&=& \left(\begin{array}{cc} \frac{\lambda_-(t) + \lambda_+(t) + (\lambda_-(t) - \lambda_+(t))\cos(4\phi_I)}{2\lambda_-(t)} & i\frac{(\lambda_-(t) - \lambda_+(t))\sin(4\phi_I)}{2\lambda_+(t)} \\ -i\frac{i(\lambda_-(t) - \lambda_+(t))\sin(4\phi_I)}{2\lambda_-(t)}  &  \frac{\lambda_-(t) + \lambda_+(t) - (\lambda_-(t) - \lambda_+(t))\cos(4\phi_I)}{2\lambda_+(t)} \end{array}\right) \,. 
\eea
Finally, we find
\bea
	F_{s}(t) &=& \frac{\lambda_-(t)^2 + 6 \lambda_-(t)\lambda_+(t) + \lambda_+(t)^2 - (\lambda_-(t) - \lambda_+(t))^2 \cos(4\phi_\nu)}{2(\lambda_-(t)\lambda_+(t) + 1)}  \,.
\eea
We see that quantum enhancement is maximized for $\phi_\nu = \pm\pi/4 +m\,\pi/2$ with integer $m$. In that case, we obtain for the QFI
\bea
	F_{s}(t) &=& \frac{(\lambda_-(t) + \lambda_+(t))^2}{\lambda_-(t)\lambda_+(t) + 1} \approx \frac{ \lambda_+(t) }{\lambda_-(t) + \lambda_+(t)^{-1}}\,,
\eea
which is {approximately the same as the} result for estimation of rotation above.

\subsection{Continuous squeezing of a squeezed probe state}
\label{sec:estcontsqueezing}

We consider the continuous sqeezing of an inital probe state by 
\begin{equation}
	\hat{H}_S =  -i \Xi \left(e^{2i\phi_{\Xi}} \hat{b}_{I_0}^{\dagger 2} - e^{-2i\phi_{\Xi}} \hat{b}_{I_0}^2 \right)/2\,,
\end{equation}
for which the time evolution of the co-rotating covariance matrix is given as
\bea
	\mbf{\Sigma}(t)_{I,\Xi} &=& \mbf{\Sigma}_{\Xi} +  e^{-\gamma_I^- t} \,\mbf{S}_\Xi(t) \cdot \left( \mbf{\Sigma}(0)_I - \mbf{\Sigma}_{\Xi}\right)  \cdot \mbf{S}_\Xi(t)
\eea
where $\mbf{\Sigma}(0)_I = \mbf{S}_I(\zeta)\mbf{S}_I(\zeta)$ and
\bea
	\mbf{S}_\Xi(t) = 
	\left(\begin{array}{cc} \cosh(\Xi t)  & -e^{2i\phi_{\Xi}} \sinh(\Xi t)   \\  
									 -e^{-2i\phi_{\Xi}} \sinh(\Xi t)  	& \cosh(\Xi t)
	\end{array}\right)\,,
\eea
and
\bea
	\mbf{\Sigma}_{\Xi} = - \frac{\gamma_I^+ }{(2\Xi)^2 - (\gamma_I^-)^2}\left(\begin{array}{cc}  
        \gamma_I^-   &  - 2 e^{2i\phi_{\Xi}} \Xi \\
	   - 2 e^{-2i\phi_{\Xi}} \Xi 	&   \gamma_I^- 
	\end{array}\right)\,.
\eea
Now, we consider the estimation of $\Xi$. For the corrsponding QFI, we find
\bea
	F_{\Xi}(t) &=& \frac{1}{2(1 + P_{\Xi}(t)^2)}\rm{Tr}\left[\left(\mbf{\Sigma}(t)_{I,\Xi}^{-1} \mbf{\Sigma}(t)_{I,\Xi}^\prime\right)^2 \right]   \,,
\eea
where
\bea
	 \mbf{\Sigma}(t)_{I,\Xi}^\prime = \left. \frac{d}{d\Xi}\mbf{\Sigma}(t)_{I,\Xi}\right|_{\Xi=0} &=&   
	 \mbf{\Sigma}_{\Xi}^\prime - e^{-\gamma_I^- t} \,\mbf{S}_\Xi(t) \cdot \mbf{\Sigma}_{\Xi}^\prime  \cdot \mbf{S}_\Xi(t) \\
	\nonumber  && + t e^{-\gamma_I^- t} \,\left[ \mbf{Y}_\Xi(t) \cdot \left( \mbf{\Sigma}(0)_I - \mbf{\Sigma}_{\Xi}\right) \cdot \mbf{S}_\Xi(t) + \mbf{S}_\Xi(t) \cdot \left( \mbf{\Sigma}(0)_I - \mbf{\Sigma}_{\Xi}\right) \cdot \mbf{Y}_\Xi(t) \right]
\eea
and
\bea
	\mbf{Y}_\Xi(t) = \left(\begin{array}{cc} \sinh(\Xi t) & -e^{2i\phi_{\Xi}}\cosh(\Xi t) \\ -e^{-2i\phi_{\Xi}}\cosh(\Xi t)  & \sinh(\Xi t) \end{array}\right)\,.
\eea
At the base point $\Xi=0$, we obtain
\bea\label{eq:derivativebasepointzero}
	 \nonumber \left. \mbf{\Sigma}(t)_{I,\Xi}^\prime \right|_{\Xi=0} &=&   
	 2\frac{\gamma_I^+ }{(\gamma_I^-)^2} \left( 1 - e^{-\gamma_I^- t} \right) \mbf{Y}_0\\
	\nonumber  && +  t e^{-\gamma_I^- t} \, \left[\mbf{Y}_0 \cdot \left( \mbf{\Sigma}(0)_I - \frac{\gamma_I^+}{\gamma_I^-}\mathbb{I} \right) +  \left( \mbf{\Sigma}(0)_I - \frac{\gamma_I^+}{\gamma_I^-}\mathbb{I}\right) \cdot \mbf{Y}_0 \right]\\
	&=& 2\frac{\gamma_I^+ }{(\gamma_I^-)^2} \left(\left( 1 - e^{-\gamma_I^- t} \right) - \gamma_I^- t\, e^{-\gamma_I^- t}\right) \mbf{Y}_0  +  t e^{-\gamma_I^- t} \, \left[\mbf{Y}_0 \cdot \mbf{\Sigma}(0)_I + \mbf{\Sigma}(0)_I  \cdot \mbf{Y}_0 \right]
\eea
where
\bea
	\mbf{Y}_0 &=&  \left(\begin{array}{cc} 0 & -e^{2i\phi_{\Xi}} \\ -e^{-2i\phi_{\Xi}}  & 0 \end{array}\right)\,.
\eea
For $\gamma_I^- t \ll 1$, we find that the first term in Eq. (\ref{eq:derivativebasepointzero}) vanishes and 
\bea
	 \left. \mbf{\Sigma}(t)_{I,\Xi}^\prime \right|_{\Xi=0} &\approx&   t \, \left[\mbf{Y}_0 \cdot \mbf{\Sigma}(0)_I + \mbf{\Sigma}(0)_I  \cdot \mbf{Y}_0 \right]\,.
\eea
Since 
\bea
	\mbf{\Sigma}(t)_{I,\Xi=0}^{-1} &=& \mbf{\Sigma}(t)_I^{-1}
\eea

Finally, we find
\bea
	F_{\Xi}(t) &=& t^2\frac{\lambda_-(t)^2 + 6 \lambda_-(t)\lambda_+(t) + \lambda_+(t)^2 - (\lambda_-(t) - \lambda_+(t))^2 \cos(4\phi_\nu)}{2(\lambda_-(t)\lambda_+(t) + 1)}  \,,
\eea
which is equivalent to the QFI for the measurement of the amplitude of instantaneous squeezing multiplied by $t^2$. It is maximized for $\phi_\nu = \pm\pi/4 +m\,\pi/2$ with integer $m$ leading to
\bea
	F_{\Xi}(t) &=& t^2\frac{(\lambda_-(t) + \lambda_+(t))^2}{(\lambda_-(t)\lambda_+(t) + 1)}  \,,
\eea
recovering the expression for instantaneous squeezing by setting $s=\Xi t$ and taking into account that $F_{\Xi}(t) =  F_{s}(t)|\frac{ds}{d\Xi}|^2$.

\section{Three-body loss in lower dimensions} 
\label{sec:lowerdim}

In the main text, we have considered only three-dimensional BECs. However,
the rate of three-body loss depends on the dimensionality of the Bose gas. 
Three-body loss in one-dimensional BECs has been investigated theoretically, 
for example, in \cite{mehta_three-body_2007,haller_three-body_2011}
and experimentally, for example, in \cite{tolra_observation_2004,haller_three-body_2011}. 
In \cite{tolra_observation_2004}, the three-body decay constant $D$ is denoted as $K_3^\rm{1D}$
(e.g. compare \cite{tolra_observation_2004} equation (1) with our equation (\ref{eq:BECdecay})).
There, 1d Bose gases of Rubidium in the Bogoliubov regime are investigated and it is
found that $K_3^\rm{1D}\approx 1.2(7)\times 10^{-30}\rm{cm}^6s^{-1}$, which 
they compare to a measurement of $K_3^\rm{3D}$ to find a 7-fold reduction of the decay
constant from 3d to 1d.
In \cite{haller_three-body_2011}, the decay constant appears as $D=3K^{(3)}g^{(3)}$ 
(see the rate equation in the third paragraph in the left column on page 2), where $K^{(3)}$
incorporates the three-body physics, e.g. the probability of dimer formation in a 
three-body scattering process, and $g^{(3)}$ is the three-body correlator of the Bose gas,
which crucially depends on the geometry of the confinement/trap.
Fig. 3 (a) of \cite{haller_three-body_2011} shows experimental values for $K^{(3)}g^{(3)}$ for Bose gases of Cs
between $10^{-30}\,\rm{cm^6 s^{-1}}$ and $10^{-29}\,\rm{cm^6 s^{-1}}$ for a wide range of scattering lengths
(tuned by a Feshbach resonance). This gives a decay constant $D$ of about the same order of magnitude as
the one we considered for Rb and Yb above. In particular, the results of \cite{haller_three-body_2011}
show at least for Cs, that the decay constants in 3d and 1d are close to each other in the low
weakly interacting Bogoliubov regime that we are considering, but the 1d decay constant may be reduced
significantly in comparison to the 3d decay constant for larger scattering lengths.

It is reasonable to assume that $K^{(3)}$ does not depend on the dimensions of the Bose gas
as long as non of the dimensions of the confinement is smaller than the scattering length or the extension of
the dimer created in the process. For Rb-87, a scattering length of $a_\rm{scatt}\approx 98 a_0$ 
has been reported in \cite{Egorov:2013meas}, where $a_0 \approx 5 \times 10^{-11}\,\rm{m}$ is the Bohr radius. 
The scattering length of Yb-168 can be found in \cite{Kitagawa:2008two} 
and leads to an interaction constant of approximately $a_\rm{scatt}\approx 250 a_0$.
Therefore, for confinement length scales of the order of $100\,\rm{nm}$ and above, $K^{(3)}$
can be assumed to be the same for Bose gases of Rb and Yb in 3d, 2d and 1d.

Theoretical predictions for $g^{(3)}$ in 1d
and its temperature dependence were given in \cite{kormos_exact_2011}.
Fig. 1 of \cite{kormos_exact_2011} shows $g^{(3)}$ for different temperatures
as a function of the dimensionless coupling constant $\gamma$. 
In the weak coupling Bogoliubov regime $\gamma \lesssim 1$, it is found
that the value of the correlator at zero temperature differs from that for large temperatures
by about one to two orders of magnitude.
When $\gamma \gtrsim \rm{max}(1,\sqrt{T/T_D})$ the Tonks-Girardeau
regime is reached, where $T$ is the temperature,
$T_D = \hbar^2 n^2/(2mk_B)$, $k_B$ is the Boltzmann constant, $m$ is the atomic mass
and $n=N_a/L$ is the 1d density of the Bose gas. In the Tonks-Girardeau regime,
$g^{(3)}$ vanishes which leads to vanishing of the decay constant, and thus, vanishing three-body loss.
However, $\gamma$ scales with the inverse of the
atom density \cite{kormos_exact_2011}, and therefore, the restriction to the Tonks-Girardeau regime leads 
to a bound on the number of atoms.

For two-dimensional weakly interacting Bose gases, it was predicted in 
\cite{Kagan:1996evo,pitaevskii_dynamics_1996,Pitaevskii:1997breath} that the 
breathing mode of the whole gas would be un-damped. A very low damping of that mode was
experimentally observed in \cite{Chevy:2002trans}.
A small damping due to vortex--anti-vortex pair creation was predicted later in \cite{Fedichev:2003zero}.

\section{Beliaev and Landau damping} 
\label{app:bellanddamp}

When the Hamiltonian $\hat{H}_0$ is expanded based on the split (\ref{eq:split}),
third and fourth order terms in the field operator $\hat{\chi}$ appear implying interactions that we neglected so far.
In particular, the third order terms lead to the damping effects called Landau damping and Beliaev damping 
that can have significant effects on the state of the field $\hat\chi$. 
In Landau damping, two quasiparticle excitations (i.e., atoms in excited motional states) interact such that their energy and momentum are combined into 
a single higher energetic quasi-particle leaving the second quasi-particle as a condensate atom. It was initially discussed in \cite{Szepfalusy:1974on,Shi:1998fin,Fedichev:1998damp}. An expression for the damping constant $\gamma_\mathbf{k}$ in a uniform BEC for general temperatures was derived in \cite{Pitaevskii:1997land,Giorgini:1998dam}.

In \cite{Pitaevskii:1997land}, an expression for the damping constant of Landau damping $\gamma_\rm{La}$ in a uniform BEC for quasi-particle energies $\hbar\omega_n\ll k_B T$ was given as
\begin{equation}\label{eq:Landaudamp0}
 	\gamma_\mathbf{k}^\rm{La} = \frac{2\sqrt{\pi} \hbar  |\mathbf{k}| a_\rm{scatt}^2\rho_0}{m} F_\rm{La}\,,
\end{equation}
where
\bea
	F_\rm{La} = 8\sqrt{\pi} \int_0^\infty dx\,(e^x-e^{-x})^{-2}\left(1-\frac{1}{2u}-\frac{1}{2u^2}\right)^2\,,
\eea
and $u=\sqrt{1+4(k_BT/\mu)^2 x^2}$. For temperatures $T$ such that $k_B T \ll \mu $, where $\mu = m c_0^2$ is the chemical potential and $c_0=1/(\sqrt{2}m\zeta)$ is the speed of sound
of the BEC, and, the Landau damping rate becomes \cite{Giorgini:1998dam}
\begin{equation}\label{eq:LandaudamplowT0}
 	\gamma_\mathbf{k}^\rm{La} = \frac{3\pi^3 }{40}\frac{|\mathbf{k}| (k_B T)^4 }{ \rho_0 \hbar^3 m c_0^4}\,.
\end{equation}
The Landau damping rate is proportional to the temperature's fourth power. 
Therefore, Landau damping can be decreased significantly by lowering the temperature further once the low temperature regime $k_B T \ll \mu $ is reached.

Beliaev damping is the scattering of a quasi-particle and a condensate atom leading to two quasi-particle excitations that share the kinetic energy 
and the momentum of the initial quasi-particle. The corresponding damping constant for uniform BECs is given as \cite{Giorgini:1998dam}
\begin{equation}\label{eq:Beliavdamp0}
	\gamma_\mathbf{k}^\rm{Be} = \gamma_\mathbf{k}^\rm{Be,0} \left[1 + 60 \int_0^1dx\,\frac{x^2(x-1)^2}{e^{\frac{\hbar \omega_\mathbf{k} }{k_BT} x}-1}\right]\,,
\end{equation}
where
\begin{equation}\label{eq:Beliavdamplow0}
	\gamma_\mathbf{k}^\rm{Be,0} = \frac{3}{640\pi}\frac{\hbar  |\mathbf{k}|^5}{m  \rho_0}\,
\end{equation} 
is the Beliaev damping constant at zero temperature. To obtain equation (\ref{eq:Beliavdamp0}),
we assumed that the quotient of the atom density of the BEC and total atom density including the thermal cloud is close to one.
This is the case for temperatures much smaller than the critical temperature of the BEC \cite{Giorgini:1997sca}.

In \cite{Howl_2017}, the effect of Landau damping and Beliaev damping has been investigated and a master equation for the quasi-particle field's
state has been derived that is very similar to the one we obtained for three-body loss here.
The difference is modified damping constants and an additional term that corresponds to the
inverse process of the second term in equation (\ref{eq:masterunif}), in which a quasi-particle in mode $n$ is created from the condensate.
For this term to be present, there has to be a bath of quasi-particles that can interact in the inverse processes of Landau and Beliaev damping.
Therefore, this is an effect of finite temperature. In particular, it leads to thermalization of the modes.
In contrast, for the three-body loss process, we assumed that molecules and highly excited atoms are removed from the system quickly
enough to consider the corresponding sectors as baths in the vacuum state, i.e. zero temperature. Therefore, the additional
term in the master equation obtained in \cite{Howl_2017} does not appear here. 

On the level of Gaussian states, combining all processes leads to a modification of the constants $\gamma^u_\mathbf{k}$ and $\gamma^v_\mathbf{k}$, where $\gamma^u_\mathbf{k}$ 
contains the spontaneous and stimulated annihilation processes and $\gamma^v_\mathbf{k}$ contains the creation processes
due to the baths. Then, the time evolution including three-body loss, Beliaev damping and Landau damping is still given by
equation (\ref{eq:oem-rotating-wave}). 
Assuming thermal baths with delta correlations for Beliaev and Landau damping, we may write 
$\gamma^u_\mathbf{k} = |\alpha_\mathbf{k}|^2\gamma_{3b} + (\bar{N}_\mathbf{k} + 1)(\gamma_\rm{L} + \gamma_\rm{B})$ and $\gamma^v_\mathbf{k} = |\beta_\mathbf{k}|^2\gamma_{3b} + \bar{N}_\mathbf{k}(\gamma_\rm{L} + \gamma_\rm{B})$, where $\bar{N}_\mathbf{k}$ is the average thermal occupation number of the mode $\mathbf{k}$ and $\gamma_\rm{L}$ and $\gamma_\rm{B}$ are the single particle (temperature-dependent) damping constants of Landau damping and Beliaev damping, respectively. 
For this case, we find 
$\gamma^-_\mathbf{k} = \gamma_{3b} + \gamma_\rm{L} + \gamma_\rm{B}$ and 
$\gamma^+_\mathbf{k}  = (2|\alpha_\mathbf{k}|^2 - 1) \gamma_{3b} + (2\bar{N}_\mathbf{k} + 1)(\gamma_\rm{L} + \gamma_\rm{B})$.
\begin{figure}[h]
\centering
\includegraphics[width=7cm,angle=0]{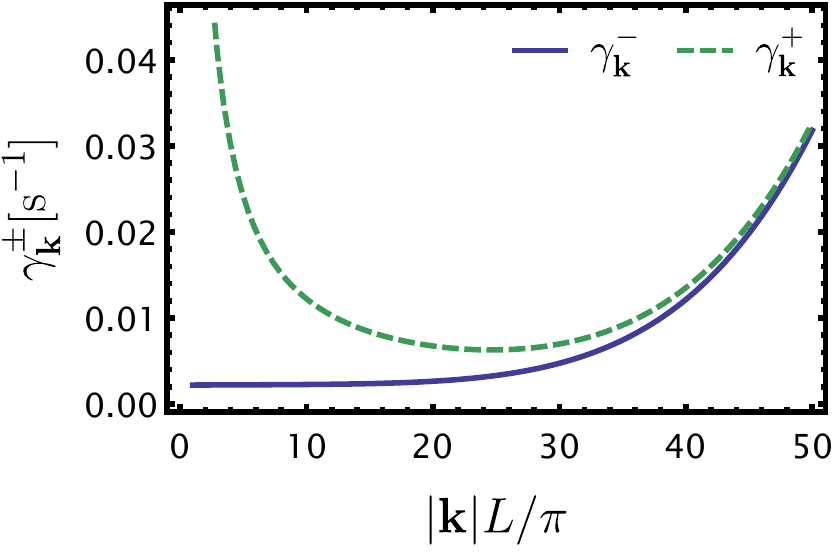}
\caption{\label{fig:fngamman} {\small Total damping constant $\gamma_\mathbf{k}^-$ (blue solid line)  and the total noise constant $\gamma_\mathbf{k}^+$ (green dashed line) are plotted as
functions of $|\mathbf{k}|L/\pi$, where $L$ is the length of the elongated direction of the cuboid BEC, based on the parameters mentioned in the text. For low momenta, the $\mathbf{k}$-independent $\gamma_{3b}$ due to three-body loss dominates. For higher quasi-particle momenta, Beliaev damping starts to dominate due to its proportionality to $|\mathbf{k}|^5$. 
In the presented regime of very low temperatures,
Landau damping is completely suppressed. The difference between $\gamma_\mathbf{k}^-$ and $\gamma_\mathbf{k}^+$ is strongly pronounced for small momenta where 
$|\alpha_\mathbf{k}|$ becomes large.  }}
\end{figure}
\newline
In the following, we will give a few numbers for the damping constants and compare to the effect of three-body loss in different regimes.
For an elongated cuboid BEC of length $L = 200\,\rm{\mu m}$ and aspect ratio $1/3$, a number of $N_a= 10^7$ atoms corresponds to a density
of $10^{13}\,\rm{cm}^{-3}$ corresponding to $\gamma_{3b}=0.2\,\rm{s}^{-1}$ and a speed of sound $c_0\approx 6\times 10^{-4}\,\rm{ms^{-1}}$
for Rb. We assume a very low temperature of $200\,\rm{pK}$. Then, we
find $\gamma_\mathbf{k}^\rm{La} \approx 5 \times 10^{-6} |\mathbf{k}|L/\pi \,\rm{s}^{-1}$ and $\gamma_\mathbf{k}^\rm{Be} \approx 1 \times 10^{-9} (|\mathbf{k}|L/\pi)^5 \,\rm{s}^{-1}$ for a BEC of Rb-atoms. This means that Landau damping is completely suppressed and Beliaev damping is suppressed for small quasi-particle momenta.
The total damping constant $\gamma_\mathbf{k}^-$ and the total noise constant $\gamma_\mathbf{k}^+$ are shown in Figure \ref{fig:fngamman}.

\end{document}